\documentclass[11pt]{article}
\usepackage{mint,psfig,mycite}

\title{
\vspace{-1cm}
\begin{flushright}
\normalsize FERMILAB-PUB-01/035-T \\
HUPD-0104 \\
RCNP-Th01005 \\
UT-CCP-P-102 \\
hep-lat/0103026
\end{flushright}
\vspace{1cm}
$O(a)$-improved quark action on anisotropic lattices
and perturbative renormalization of heavy-light currents}

\author{
Junpei Harada$^a$, Andreas S.~Kronfeld$^{b,c}$, Hideo Matsufuru$^d$, \\
Noriaki Nakajima$^d$ and  Tetsuya Onogi$^a$\\[1.2em]
	{\it $^a$Department of Physics, Hiroshima University,
		Higashi-Hiroshima 739-8526, Japan} \\
	{\it $^b$Theoretical Physics Department, Fermi National Accelerator
		Laboratory,}\\
		{\it Batavia, Illinois 60510, U.S.A.}\\
	{\it $^c$Center for Computational Physics, University of Tsukuba,}\\
		{\it Tsukuba 305-8577, Japan}\\
	{\it $^d$Research Center for Nuclear Physics, Osaka University,
		Ibaraki 567-0047, Japan} }

\date{March 22, 2001}

\begin{document}

\maketitle

\begin{abstract}
We investigate the Symanzik improvement of the Wilson quark action on
anisotropic lattices.
Taking first a general action with nearest-neighbor and clover 
interactions, we study the mass dependence of the ratio of the hopping 
parameters, the clover coefficients, and an improvement coefficient for 
heavy-light vector and axial vector currents.
We show how tree-level improvement can be achieved.
For a particular choice of the spatial Wilson coupling, the results
simplify, and $O(m_0a_\tau)$ improvement is possible.
(Here $m_0$ is the bare quark mass and $a_\tau$ the temporal lattice
spacing.)
With this choice we calculate the renormalization factors of  
heavy-light bilinear operators at one-loop order of perturbation 
theory employing the standard plaquette gauge action.
\end{abstract}

\section{Introduction}
\label{sec:Introduction}

The anisotropic lattice has become an important tool in
lattice QCD simulations.
With a small temporal lattice spacing~$a_\tau$ one can more easily
follow the time evolution of correlators, while keeping the spatial
lattice spacing~$a_\sigma$ comparatively modest~\cite{Kar82}.
This approach is especially effective when the signal-to-noise ratio
deteriorates quickly, as, for example, in the case of
glueballs~\cite{Morningstar:1997ff}.
The better signal-to-noise ratio is beneficial also for heavy quark
systems~\cite{Fingberg:1998qd}.
In addition,
it is hoped that the anisotropy can be exploited to reduce lattice
artifacts~\cite{Kla99}, which are a special concern with heavy quarks.

In current work on heavy quarks, lattice artifacts are controlled with
non-relativistic~QCD (NRQCD) and heavy-quark effective theory (HQET).
This is done either \emph{a priori}, by discretizing the NRQCD 
action~\cite{Lepage:1987gg}, or \emph{a posteriori}, by using the 
effective theories to describe lattice gauge theory with Wilson 
fermions~\cite{EKM97,Kronfeld:2000ck}.
These strategies are possible because the typical \emph{spatial}
momenta in heavy quark systems are much smaller than the heavy quark
mass.
Heavy quarkonia have momenta $\vec{p}\sim m_Qv$ and $k\sim m_Qv^2$,
where $v\sim0.1$--0.3 is the heavy-quark velocity; heavy-light
hadrons have momenta only of order~$\Lambda_{\rm QCD}$.
In these approaches one is left with discretization effects of order
$(\Lambda_{\rm QCD}a)^n$ from the light quarks and gluons and of order
$\alpha_s^l(\vec{p}/m_Q)^n$ from the heavy quark.

The method of Ref.~\cite{EKM97} smoothly connects to the usual
continuum limit, so one can, in principle, reduce discretization
effects to scale as a power of the lattice spacing~$a$, but only by
making~$a$ too small to be practical.
Klassen proposed using anisotropic lattices with the anisotropy
$\xi=a_\sigma/a_\tau$ chosen so that $m_Qa_\tau$ and $\vec{p}a_\sigma$
are both small~\cite{Kla99}.
Clearly, this proposal works only if $\vec{p}$ is smaller than $m_Q$,
as in the approaches based explicitly on heavy-quark theory.
It also works only if renormalization constants have a smooth limit as
$m_0a_\tau\to 0$, where $m_0$ is the bare quark mass.
In particular, one would like to be able to expand the renormalization
constants in powers of $m_0a_\tau$ even when $m_0a_\sigma\sim1$.
Then it may be possible to adjust the improvement parameters of the 
lattice action (and currents) in a non-perturbative, mass-independent 
scheme~\cite{Luscher,Kla99}.
If, on the other hand, $m_0 a_{\sigma}$ dependence appears in an
essential way, then one would be forced back to a non-relativistic
interpretation, as explained for isotropic lattices in
Refs.~\cite{EKM97,Kronfeld:2000ck}.

To our knowledge there is no proof that cutoff effects always appear
as powers of~$m_0a_{\tau}$.
In this paper we try to gain some experience by calculating the full 
mass dependence of several (re)normalization constants, first at tree 
level and then at one-loop in perturbation theory.
We focus on the Fermilab action~\cite{EKM97}, which is the most general action
without doubler states, having different nearest-neighbor and clover
couplings in the temporal and spatial directions.
This action has been applied on anisotropic lattices to the charmonium
system~\cite{Kla99,TARO00,Ume00,Chen00,CPPACS00}, as have some actions
with next-to-nearest neighbor interactions.
The self-energy has been calculated at the one-loop level in
perturbation theory~\cite{GS00}.

In the numerical work on charmonium, two different choices for tuning
the spatial Wilson term have been made.
One choice is that of Refs.~\cite{TARO00,Ume00}, where $r_s=1/\xi$.
Another choice is that of Refs.~\cite{Kla99,Chen00,CPPACS00},
where $r_s=1$.
In the first part of this paper, we study improvement conditions for
these two choices, as a function of the heavy quark mass.
(In a perturbative calculation more generally improved actions with 
$r_t=\xi^2\zeta r_s$ also have been considered~\cite{GS00}.)
By studying the full functional dependence on $\xi$ and $m_0a_\tau$,
we can test whether $m_0a_\sigma$ appears in an essential way.
We find that the limit of small $m_0a_\tau$ is benign at the tree-level
only for the first choice, $r_s=1/\xi$.
For the other choice, $r_s=1$, the continuum limit is reached only for
$m_0a_\sigma\ll1$.

It turns out that with the first choice ($r_s=1/\xi$) two of the
improvement parameters vanish at the tree-level as $m_0a_\tau\to 0$.
This simplifies the one-loop analysis, so in the second part of the
paper we concentrate on this choice.
This calculation has two purposes.
The first is to study cutoff effects of the renormalization
coefficients and to test at the one-loop level whether they still
appear only as powers of $m_0 a_{\tau}$.
The second is for phenomenological applications to heavy-light matrix
elements.
Even if a non-relativistic interpretation is necessary, anisotropic
lattices are a good method for reducing the signal-to-noise
ratio~\cite{Fingberg:1998qd}.

This paper is organized as follows.
Sec.~\ref{sec:Formulation} describes the quark field action and
discusses its parametrization in detail.
In Sec.~\ref{sec:One-loop}, the expression for the one-loop
perturbative calculation is given.
The numerical result for these perturbative constants are presented in
Sec.~\ref{sec:Result}.
The last section is devoted to our conclusions.
We give the Feynman rules in Appendix~\ref{sec:Feynman} and explicit
expressions for the one-loop diagrams in
Appendix~\ref{sec:expressions}.

\section{Anisotropic quark action}
\label{sec:Formulation}

This section describes the actions with Wilson 
fermions~\cite{Wilson:1977nj} on anisotropic lattices.
We denote the renormalized anisotropy with~$\xi$, and the spatial and
temporal lattice spacings with~$a_{\sigma}$ and~$a_{\tau}$
respectively:
\begin{equation}
	a_{\sigma} = \xi a_{\tau}.
\end{equation}
These lattice spacings would be defined through the gauge field with
quantities such as the Wilson loops or the static quark potential.
We therefore consider $\xi$ to be independent of the quark mass.

\subsection{Quark field action}
\label{sec:qaction}

Following Ref.~\cite{EKM97}, let us introduce an action with two 
hopping parameters~\cite{Wilson:1977nj} and two clover~\cite{SW85} 
coefficients,
\begin{eqnarray}
S &=& \sum_n \bar{\psi}_n 
 \Big[  \psi_n  - \kappa_t   [( r_t -\gamma_4) U_{n,4} \psi_{n+\hat{4}}  
      + (r_t +\gamma_4) U^{\dagger}_{n-\hat{4},4} \psi_{n-\hat{4}} ]
	   \nonumber \\
& & - \kappa_s \sum_i 
    [ (r_s - \gamma_i) U_{n,i} \psi_{n+\hat{\imath}}  
    + (r_s + \gamma_i) U^\dagger_{n-\hat{\imath},i}\psi_{n-\hat{\imath}}]
	  \label{eq:kappa_action} \\
& & + \ihalf  r_s c_B \kappa_s \sum_{i,j,k} \epsilon_{ijk} 
       \sigma_{ij} B_{n,k} \psi_n 
       + i c_E \kappa_s \sum_{i} \sigma_{4i} E_{n,i} \psi_n \Big].
	   \nonumber
\end{eqnarray}
This is the most general nearest-neighbor clover action.
Note that the notation is slightly different than in
Ref.~\cite{EKM97}; $c_B$ of Ref.~\cite{EKM97} corresponds to $r_sc_B$
in~(\ref{eq:kappa_action}).

It is helpful to change to a notation with a quark mass.
We rescale field by
\begin{equation}
	\psi_n = \frac{a_{\sigma}^{3/2}}{\sqrt{2\kappa_s}} \psi(x),
\end{equation}
and introduce the bare mass
\begin{equation}
	m_0 a_{\tau} = \frac{1}{2\kappa_t} - [r_t+3 r_s\zeta],
\end{equation}
with $\zeta=\kappa_s/\kappa_t$.
Then one can rewrite the action as
\begin{eqnarray}
	S &=& a_{\tau}a_{\sigma}^3\sum_x \bar{\psi}(x) \left[
		m_0 +
		\half (r_t+\gamma_4) D_4^{-} -
		\half (r_t-\gamma_4) D_4^{+} +
		\xi \zeta  \vecprd{\gamma}{D} -
		\half  a_{\tau} \xi^2 r_s      \zeta \triangle^{(3)} \right.
	  \nonumber \\
	  & & \left. - 
	  	\ihalf a_{\tau} \xi^2 r_s  c_B \zeta \vecprd{\Sigma}{B} -
		\half  a_{\tau} \xi        c_E \zeta \vecprd{\alpha}{E} 
		\right] \psi(x).
	\label{eq:mass_action}
\end{eqnarray}
The covariant difference operators~$D_4^\mp$, $\vec{D}$
and~$\triangle^{(3)}$, and the fields $\vec{B}$ and $\vec{E}$ are
defined as in Ref.~\cite{EKM97}, except that the lattice spacing~$a$ is
replaced by~$a_{\tau}$ or~$a_{\sigma}$ in the obvious way.

The action~$S$ has six parameters $m_0$, $r_t$, $r_s$, $\zeta$, $c_B$,
and~$c_E$.
Two are redundant and can be chosen to solve the doubling
problem~\cite{EKM97}.
In particular, we choose $r_t=1$ to eliminate doubler states.
We then rename $r_s=r$, but discuss how to adjust it below.
The other four parameters are dictated by physics.
The bare mass is adjusted to give the desired physical quark mass,
and $\zeta$, $c_B$, and $c_E$ are chosen to improve the action.

Following Ref.~\cite{EKM97} we also consider a rotated field
\begin{equation}
	\Psi(x) = e^{M_1 a_{\tau}/2}
		\left[ 1 + a_{\sigma} d_1 \vecprd{\gamma}{D}\right]\psi(x),
	\label{eq:rot}
\end{equation}
where $M_1$ is the rest mass, defined and given below, and
$d_1$ is an improvement parameter.
This field is convenient for constructing heavy-light bilinears
\begin{eqnarray}
	V_\mu^{ub} & = & \bar{\psi}^u\gamma_\mu\Psi^b , \\
	A_\mu^{ub} & = & \bar{\psi}^u\gamma_\mu\gamma_5\Psi^b ,
\end{eqnarray}
which, at the tree-level, are correctly normalized currents
for all~$m_0a_\tau$.
Beyond the tree-level one may add dimension-four terms to these
currents, and one must multiply with suitable renormalization factors.

The renormalization factors and the improvement parameters $\zeta$, 
$c_B$, $c_E$, and~$d_1$ must, in general, be chosen to be functions 
of~$m_0a_\tau$ and the anisotropy~$\xi$.
Below we shall give the full mass dependence to check whether,
for small~$m_0a_\tau$, power series such as
\begin{equation}
	\zeta(\xi,m_0a_\tau) = \zeta(\xi,0) + m_0a_\tau \zeta'(\xi,0)
	\label{m0at-exp}
\end{equation}
can be admitted.
If $m_0a_\sigma=\xi m_0a_\tau$ enters into the full mass-dependent
expression, this series would not be accurate when $m_0a_\sigma\sim1$.
In the past~\cite{Kla99} the behavior in~(\ref{m0at-exp}) was
implicitly assumed.
If the expansions of the form~(\ref{m0at-exp}) do work, then for full
$O(a)$ improvement one must adjust $\zeta(\xi,0)$, $\zeta'(\xi,0)$,
$c_B(\xi,0)$, and $c_E(\xi,0)$ in the action, and
$Z_{J_{\Gamma}}(\xi,0)$, $Z'_{J_{\Gamma}}(\xi,0)$, and
$d_1(\xi,0)$ of the currents~$J_{\Gamma}=V_\mu$,~$A_\mu$.

From~(\ref{eq:mass_action}) one can see that conditions for the
improvement coefficients can be obtained by simply replacing
\begin{eqnarray}
	\zeta &\rightarrow& \xi \zeta , \\
	 r_s  &\rightarrow& \xi r , \\
	 c_B  &\rightarrow& \xi r  c_B , \\
	 c_E  &\rightarrow& c_E ,
\end{eqnarray}
in formulae in Ref.~\cite{EKM97}.
For example, the energy of a quark with momentum $\vec{p}$ is
given by
\begin{equation}
	\cosh Ea_\tau =  1 +
		\frac{(m_0a_\tau+\half\xi^2r\zeta\hat{\vec{p}}^2a_\tau^2)^2+
			\xi^2\zeta^2\vec{S}^2a_\tau^2}%
		{2(1+m_0a_\tau+\half\xi^2r\zeta\hat{\vec{p}}^2a_\tau^2)},
	\label{coshE}
\end{equation}
where $\hat{p}_i=2a_\sigma^{-1}\sin(\half p_ia_\sigma)$
and   $S_i=a_\sigma^{-1}\sin(p_ia_\sigma)$.
For small momentum $E^2=M_1^2+\vec{p}^2M_1/M_2+ O(\vec{p}^4a^2)$,
where the rest mass~$M_1$ and kinetic mass~$M_2$ are
\begin{eqnarray}
	M_1 a_{\tau} &= & \ln(1+m_0 a_{\tau}) ,  \label{M1} \\
	\frac{1}{M_2 a_{\tau}} &= & \xi^2 \left(
		\frac{2\zeta^2}{m_0 a_{\tau} (2+m_0 a_{\tau})} +
		\frac{r \zeta}{1+m_0 a_{\tau}}
		\right) . \label{M2}
\end{eqnarray}
To obtain a relativistic quark one sets the rest mass and kinetic
mass equal to each other.
This yields the condition 
\begin{equation}
	\xi\zeta = \sqrt{ \left(
		\frac{\xi r m_0 a_{\tau}(2+m_0 a_{\tau})}{4(1+m_0 a_{\tau})}\right)^2
		+\frac{m_0 a_{\tau} (2+m_0 a_{\tau})}{2 \ln(1+m_0 a_{\tau})} }
		-\frac{\xi r m_0 a_{\tau}(2+m_0 a_{\tau})}{4(1+m_0 a_{\tau})} ,
	\label{zeta}
\end{equation}
which can be read off from Ref.~\cite{EKM97}. 
Matching of on-shell three-point functions yields the conditions
\begin{eqnarray}
	c_B & = & 1, \label{cB} \\
	c_E & = & \frac{(\xi\zeta)^2 -1}{m_0 a_{\tau}(2+m_0a_{\tau})}
         + \frac{\xi^2 r \zeta}{1+m_0 a_{\tau}}
         + \frac{\xi^2 r^2m_0a_\tau(2+m_0a_\tau)}{4(1+m_0 a_{\tau})^2}
	\label{cE}
\end{eqnarray}
on the clover coefficients, and
\begin{eqnarray}
	d_1 &=& \frac{\xi \zeta (1+m_0a_\tau)}{m_0a_\tau(2+m_0a_\tau)}
        - \frac{1}{2 M_2 a_\tau} \label{d1M2} \\
    &=& \frac{\xi \zeta [2(1+m_0a_\tau)^2 - \xi rm_0a_\tau(2+m_0a_\tau) 
              - 2\xi\zeta(1+m_0a_\tau)]}
             {2m_0a_\tau(2+m_0a_\tau)(1+m_0a_\tau)} \label{d1}
\end{eqnarray}
on the rotation parameter.
These tree-level formulae~(\ref{coshE})--(\ref{d1}) have been
obtained independently by M.~Okamoto~\cite{Okamoto}.
We see that essential dependence on $m_0a_\sigma=\xi m_0a_\tau$ indeed
may arise, depending on how $r$ is tuned.

From the energy~(\ref{coshE}) one can also find the energy of states at
the edge of the Brillouin zone.
The energy of a state with $n$ components of $\vec{p}$ equal to
$\pi/a_\sigma$ is
\begin{equation}
	E_na_\tau = \ln(1+m_0a_\tau+2nr\zeta).
\end{equation}
Although there is some freedom to choose~$r$, discussed below,
one still wants to keep $E_n$ and $M_1$ well separated.

For small~$m_0a_\tau$ the interesting Taylor expansions are
\begin{eqnarray}
	\zeta & = & \xi^{-1} + \half (\xi^{-1}- r) m_0 a_{\tau} 
		+ O((m_0 a_{\tau})^2 ),
		\label{ii-zeta} \\
	 c_E & = & \half(1+\xi r) + O(m_0 a_{\tau}), \\
	 d_1 & = & \quarter(1-\xi r)
		+ O(m_0 a_{\tau}).
\end{eqnarray}
With the mass dependent factor in~(\ref{eq:rot}) there is no mass
dependence at the tree-level in the currents' normalization factors.

Let us now discuss the choice of the redundant coefficient of the
spatial Wilson term~$r$.
Two choices have been used in numerical calculations:
\begin{enumerate}
	\item[(i)]  $r=1/\xi$ \quad  \cite{TARO00,Ume00}.
		This is a natural choice because then the mass form of the
		action takes a symmetric-looking form, without $\xi$.
		In the small~$m_0a_\tau$ limit, the tree-level improvement 
		parameters become
		\begin{eqnarray}
		\zeta(\xi,0)  &=& \xi^{-1} \\
		\zeta'(\xi,0) &=& 0 \\
		c_B(\xi,0)    &=& 1 \\
		c_E(\xi,0)    &=& 1 \\
		d_1(\xi,0)    &=& 0
		\end{eqnarray}
		A key advantage is that $m_0a_\sigma$ does not appear;
		the continuum limit is reached for small~$m_0a_\tau$.
		Furthermore, both $\zeta'(\xi,0)$ and $d_1(\xi,0)$ vanish at
		the tree-level,
		which is especially helpful in one-loop calculations.
		A disadvantage is that with $r=\zeta=1/\xi$ the energy splitting
		between physical states and states at the edge of the Brillouin
		zone is not large.
		One can circumvent this problem to some extent by choosing
		appropriate cutoffs and anisotropy~\cite{Ume00}.
	\item[(ii)] $r=1$ \quad \cite{Kla99,Chen00,CPPACS00}.
		Now all hopping terms in~(\ref{eq:kappa_action}) have
		projection matrices $\half(1\mp\gamma_\mu)$, and the
		anisotropic nature appears only in~$\zeta$ and $c_B\neq c_E$.
		But now, if one considers what happens to the conditions when
		$m_0a_\sigma\ltap 1$ while $m_0a_\tau\ll 1$, then
		\begin{eqnarray}
		\xi\zeta(\xi,m_0a_\tau)  &=& 
			1 - \half m_0a_\sigma + \eighth(m_0a_\sigma)^2 +
			\half m_0a_\tau\left[1 - \half m_0a_\sigma\right] \\
		c_E(\xi,m_0a_\tau)    &=& \half\left[1 +
			\xi\left(1 + \half m_0a_\sigma +
			{\textstyle\frac{1}{8}}(m_0a_\sigma)^2\right) \right] \\
		d_1(\xi,m_0a_\tau)    &=&
			\frac{(m_0a_\sigma)^2}{16m_0a_\tau} + \quarter(1-\xi)
		\end{eqnarray}
		keeping terms of order $(m_0a_\sigma)^2$
		and~$m_0a_\tau m_0a_\sigma$.
		Clearly, the continuum limit sets in only when $m_0a_\sigma\ll1$.
		Even then $\zeta'$ and $d_1$ are non-zero
		already at the tree-level.
		An~advantage is that the splitting between the physical
		states and the edge of the Brillouin zone is larger than in
		case~(i).

		In passing we mention that Refs.~\cite{Kla99,Chen00,CPPACS00}
		take
		\begin{equation}
			\zeta = \frac{1}{\xi}\frac{2+m_0a_\tau}{2+\xi m_0a_\tau}
		\end{equation}
		which agrees with the Taylor expansion~(\ref{ii-zeta}), but
		does not agree with the full mass dependence~(\ref{zeta}).
		The denominator of this expression also is of the form
		that reaches the continuum limit only when
		$m_0a_\sigma\ll 1$.
\end{enumerate}

\begin{figure}[bp]
	\centerline{
	\leavevmode\psfig{file=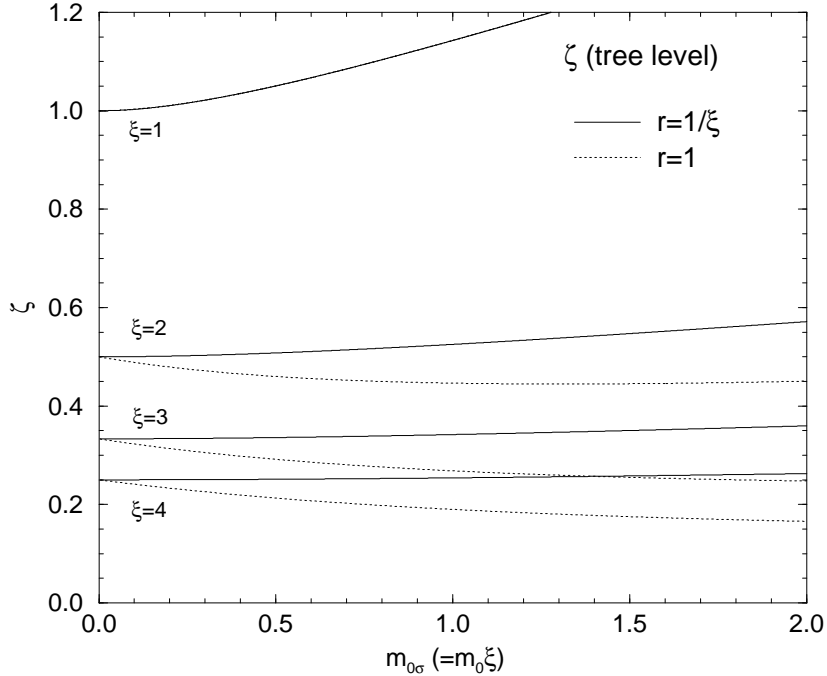,width=\figwidth}}
	\caption{The tree-level relation of $\zeta$ with the quark mass in
		the spatial lattice unit, $m_{0\sigma}$, for various $\xi$.
		The solid and the dotted lines are for choices~(i)
		and~(ii) respectively.}
	\label{fig:zeta_tree}
\end{figure}

It is instructive to examine the difference between the two conditions
on $\zeta'$ by considering the full mass dependence of~$\zeta$.
Figure~\ref{fig:zeta_tree} plots the right-hand side of~(\ref{zeta})
against $m_{0\sigma}:=m_0a_\sigma$, for several values of $\xi$ and the
two choices $r=1/\xi$ and $r=1$.
The mass in spatial lattice units, $m_{0\sigma}$ is chosen not because
it is a natural variable, but because one usually would first fix
the spatial lattice spacing $a_\sigma$ so that $\vec{p}a_\sigma$ is
small enough, while $m_0a_\sigma\sim1$.
One would then choose the anisotropy~$\xi$ to make $m_0a_\tau$ small.
For example, let us consider the charmed quark on a lattice with
$a_{\sigma}^{-1}= 1$~GeV.
The quark mass in spatial lattice units is $m_{0\sigma}\simeq 1.2$,
so if $\xi=4$, then $m_0a_\tau\simeq0.3$, which seems small.
For $r=1/\xi$ one finds
$\zeta(\xi,m_0a_\tau)=\zeta(4,0.3)\approx 0.26$,
which is only 4 percent larger than $\zeta(4,0)=0.25$.
In this sense, $m_0a_\tau\simeq0.3$ \emph{is} small.
On the other hand, for the choice $r=1$, 
$\zeta(\xi,m_0a_\tau)=\zeta(4,0.3)\approx 0.20$,
which is 20 percent smaller than $\zeta(4,0)$.
Even worse, the Taylor expansion~(\ref{ii-zeta}) estimates only~0.14.

Thus, only with the choice $r=1/\xi$ does it seem possible to
approximate the improvement coefficients by the small~$m_0a_\tau$
limit.
With this choice it seems possible to treat charmed quarks, without
appealing to the heavy-quark expansion, at accessible spatial lattice
spacings and anisotropy around~3--4.
Lattice artifacts appear under control and there is probably enough
room between $M_1$ and the energies at the edge of the Brillouin zone,
$E_n$, to accommodate the lowest excitations of the $D$~meson.
On the other hand, it seems that reasonable choices of
$a_\sigma$ and $\xi$ do not exist for treating the $b$ quark:
$m_ba_\tau$ remains big, requiring a non-relativistic interpretation
along the lines of Refs.~\cite{EKM97,Kronfeld:2000ck}.

The choice $r=1/\xi$ requires no tree-level rotation for the quark
field.
This is a great simplification for one-loop renormalization.
Then the quark and anti-quark field operators are multiplied by the
factor $\exp(M_1a_\tau/2)=1+\half M_1a_\tau + O( (M_1a_\tau)^2 )$.
With the choice $r=1$ one would have to include the rotation term for a
consistent $O(a_\tau)$ calculation.
In the rest of this paper, we therefore focus on $r=1/\xi$.

\section{One-loop Renormalization}
\label{sec:One-loop}

To carry out one-loop perturbative calculations, we must specify the
gauge field action as well as the quark action.
We begin this section with the gauge field action and remark on the
gauge couplings, which, in general, differ for the spatial and
temporal components of the gauge field.
Feynman rules required at one-loop level are summarized
in Appendix~\ref{sec:Feynman}.

\begin{figure}[bp]
\begin{center}
\leavevmode\psfig{file=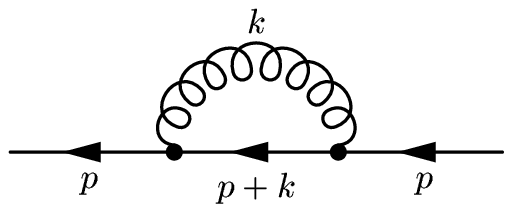,width=0.45\figwidthb}
 \hspace{0.5cm}
\leavevmode\psfig{file=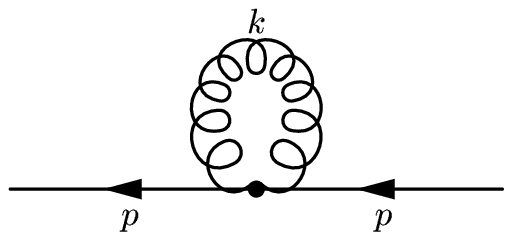,width=0.45\figwidthb}
 \hspace{0.5cm}
\leavevmode\psfig{file=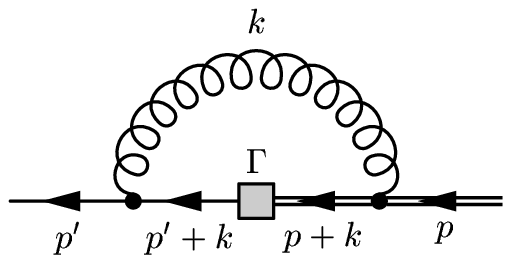,width=0.45\figwidthb} 
\vspace{-0.6cm} \\
\hspace*{-0.3cm} (a) \hspace{4.2cm} (b) \hspace{4.2cm} (c)
\label{fig:diagram}
\caption{Feynman diagrams for the quark self-energy~(a)
and~(b), and for the vertex correction~(c). }
\end{center}
\end{figure}

The self-energy at the one-loop level is represented by
the diagrams in Fig.~\ref{fig:diagram}(a)--(b).
We calculate, as a function of $m_0a_\tau$, the one-loop contribution
to the quark rest mass and wave function renormalization factor.
These quantities require the self-energy and its first derivative with
respect to the external momentum~$p_4$, evaluated on the mass shell
$(p_4,\vec{p})=(iM_1, \veg{0})$~\cite{MKE98}.
By obtaining the full mass dependence, we can check how the one-loop
corrections behave for $m_0a_\sigma\sim1$ and $m_0a_\tau$ small.
We also discuss mean field improvement of the self-energy.
In the past, the full mass dependence of the one-loop quark self-energy
has been obtained
for the Wilson action on isotropic lattices~\cite{Kuramashi:1998tt}
for the Fermilab action on isotropic lattices~\cite{MKE98}, and
for several improved actions with $r_t=\xi^2\zeta r_s$ on
anisotropic lattices~\cite{GS00}.

We also discuss vertex corrections at the one-loop level, shown in
Fig.~\ref{fig:diagram}(c), and present matching factors for the vector
and axial vector currents.
We again obtain the full mass dependence first, and use it to study the
practical situation with $m_0a_\sigma\sim1$ and $m_0a_\tau$ small.
In the past, the full mass dependence of the one-loop quark vertex
functions has been obtained on isotropic lattices for
the Wilson action~\cite{Kuramashi:1998tt},
the clover action~\cite{Ishikawa:1997xh}, and
the Fermilab action~\cite{Harada}.

\subsection{Gauge field action}

We use the standard Wilson gauge action on the anisotropic
lattice~\cite{Kar82}:
\begin{equation}
	S_{\rm gauge} = \beta a_\tau a_\sigma^3 \sum_x \left[
		\sum_{i<j=1}^3 \frac{1}{\gamma_G}
			\left(1 - \third \Re \tr U_{ij}(x) \right) +
		\sum_{i=1}^3   \gamma_G
			\left(1 - \third \Re \tr U_{i4}(x) \right) \right],
	\label{eq:Sgauge}
\end{equation}
where $U_{\mu\nu}$ denotes parallel transport around a plaquette in the
$\mu\nu$~plane.
The bare anisotropy $\gamma_G$ coincides with the renormalized
isotropy~$\xi$ at the tree-level.
In gauge field theory with $N_c$~colors, the coupling~$\beta$ is
related to the usual bare gauge coupling~$g_0$ by $\beta=2N_c/g_0^2$.

There is a subtlety in the gauge coupling, because the temporal and 
spatial gluons have different couplings~\cite{Kar82}.
One can rewrite $\beta$ and $\gamma_G$ as
\begin{eqnarray}
	\beta_{\sigma} &=& \frac{2N_c}{g^2_{\sigma}(a_{\sigma},\xi)}
		=  \frac{\beta}{\gamma_G} , \\
	\beta_{\tau}   &=& \frac{2N_c}{g^2_{\tau}(a_{\sigma},\xi)}
		=  \beta \gamma_G,
\end{eqnarray}
where $g_\sigma^2$ and $g_\tau^2$ are couplings for spatial and temporal
gluons, respectively.
Although at the one-loop level $g_{\sigma}^2$ and $g_{\tau}^2$ need not
be distinguished, it is convenient to separate the results for spatial
and temporal parts.
To improve perturbative series, it is crucial to use renormalized
couplings, defined at the momentum scale typical for the process under
consideration~\cite{LM93}.
These couplings, and therefore the scales, could be defined separately
for spatial and temporal gluons.
With this end in mind, we shall present the coefficients 
of~$g_{\sigma}^2$ and~$g_{\tau}^2$ separately.

\subsection{Rest mass renormalization}

The relation between the rest mass to the self-energy is~\cite{MKE98}
\begin{equation}
	e^{M_1a_\tau} = 1 + m_0a_\tau - \tr[P_+\Sigma(iM_1,\veg{0})]a_\tau,
	\label{rest mass}
\end{equation}
where $P_+=(1+\gamma_4)/2$ and the self-energy $\Sigma(p_4,\vec{p})$ is
the sum of all one-particle irreducible two-point diagrams.
The formula~(\ref{rest mass}) is valid for all masses and at every
order in perturbation theory.
Since it is obtained from the pole position, the rest mass is infrared 
finite and gauge independent at every order in perturbation 
theory~\cite{Kronfeld:1998di}.
We write the perturbation series as
\begin{equation}
	\Sigma(ip_4, \vec{p}) =
		\sum_{l=1}^\infty g^{2l} \Sigma^{[l]}(p_4,\vec{p}; m_0),
\end{equation}
where we explicitly specify the bare quark mass~$m_0$.
The quark is massless ($M_1=0$) when the bare mass is tuned to
\begin{equation}
	m_{0c} = \tr[P_+\Sigma(0,\veg{0};m_{0c})].
\end{equation}
It is more convenient to introduce a subtracted bare mass
$M_0=m_0-m_{0c}$, which vanishes for a massless quark.
Then the formula for the rest mass becomes
\begin{equation}
	e^{M_1a_\tau} = 1 + M_0a_\tau -
		\tr[P_+\bar{\Sigma}(iM_1, \veg{0}; M_0)]a_\tau,
	\label{M1renorm}
\end{equation}
where
\begin{equation}
	\bar{\Sigma}(p_4,\vec{p};M_0)] = \Sigma(p_4,\vec{p};m_0) - m_{0c}.
\end{equation}
In developing the perturbation series, now $M_0$ is treated
independently from~$g^2$.

The perturbative series for~$M_1$ is
\begin{equation}
	M_1 = M_1^{[0]} + \sum_{l=1}^\infty g^{2l} M_1^{[l]} .
\end{equation}
From~(\ref{M1renorm})
\begin{eqnarray}
	M_1^{[0]} & = & a_\tau^{-1} \ln (1+M_0a_\tau) ,   \\
	M_1^{[1]} & = & -
		\frac{\tr[P_+\bar{\Sigma}^{[1]}(iM_1,\veg{0};M_0)]}%
		{1+M_0a_\tau}.
\label{eq:mass_renorm2}
\end{eqnarray}
In evaluating $\Sigma^{[1]}$ one may disregard the distinction between
$M_0$ and~$m_0$, because $m_{0c}$ starts at one-loop order.
To show the mass dependence it is convenient~\cite{MKE98} to introduce
the multiplicative renormalization factor~$Z_{M_1}$ defined by,
\begin{equation}
	M_1a_\tau = Z_{M_1} \tanh M_1^{[0]}a_\tau.
\end{equation}
From $Z_{M_1}^{[1]}=M_1^{[1]}a_\tau/\tanh M_1^{[0]}a_\tau$ one can then
remove the anomalous dimension by writing
\begin{equation}
	Z_{M_1}^{[1]} = C_F \left[c^{[1]} -
		\frac{3}{16\pi^2}\ln(M_1^2 a_{\tau}^2)\right].
\end{equation}
Numerical results for~$M_1^{[1]}a_\tau$ and~$c^{[1]}$ are given in
Sec.~\ref{sec:Result}.

\subsection{Wave function renormalization}

The all orders formula for the wave function renormalization 
factor is~\cite{MKE98}
\begin{equation}
	Z_2^{-1} = e^{M_1a_\tau} - \tr[P_+\dot{\Sigma}(iM_1,\veg{0};M_0)]a_\tau
	\label{Z2}
\end{equation}
where
\begin{equation}
	\dot{\Sigma}(p_4,\vec{p};M_0) = \frac{1}{i}
		\frac{\partial\bar{\Sigma}}{\partial p_4}(p_4,\vec{p};M_0).
\end{equation}
In view of the mass dependence, we write
\begin{equation}
	(1+M_0a_\tau)Z_2 = 1 + \sum_{l=1}^\infty g^{2l} Z_2^{[l]},
\end{equation}
so that the $Z_2^{[l]}$ are only mildly mass dependent.
This definition of~$Z_2^{[l]}$ is slightly different from that of 
Ref.~\cite{MKE98} for $l>1$.

The wave function renormalization factor is infrared divergent and 
gauge dependent.
Therefore we express the one-loop term as
\begin{eqnarray}
	Z_2^{[1]} & = & (1+M_0a_{\tau})^{-1} 
		\tr[P_+(\bar{\Sigma}^{[1]}+\dot{\Sigma}^{[1]})], \\
		& =: & W^{[1]} + L^{[1]},
 \label{eq:WL_latt}
\end{eqnarray}
where $W^{[1]}$ and $L^{[1]}$ are the infrared finite and singular 
parts, respectively.
The infrared divergence does not depend on the ultraviolet regulator, 
so it is the same as in the continuum theory.
We define $L^{[1]}$ by the continuum expression.
For a massive quark ($\lambda^2\ll m^2\ll a_\tau^{-2}$) in Feynman 
gauge,
\begin{equation}
	 L_{h}^{[1]} = \frac{C_F}{16\pi^2} \left[- \frac{9}{2} + 
	 	3 \ln(m^2 a_{\tau}^2) - 2 \ln (\lambda^2a_{\tau}^2)\right],
 \label{eq:Z2h_latt}
\end{equation}
which we use for the heavy quark.
In particular, $W_h^{[1]}$ is defined by combining~(\ref{eq:WL_latt})
and~(\ref{eq:Z2h_latt}) with $m=M_2^{[0]}$.
For a massless quark ($m^2=0$, $\lambda^2\ll a_\tau^{-2}$), the mass 
singularity seen in~(\ref{eq:Z2h_latt}) can still be regulated by the
gluon mass.
In Feynman gauge,
\begin{equation}
 	 L_l^{[1]} = \frac{C_F}{16\pi^2} \ln (a_{\tau}^2\lambda^2),
	\label{eq:Z2l_latt}
\end{equation}
which we use for the light quark.
Here $C_F=N_c^2 - 1/2N_c$ [$=4/3$ for SU(3)].
Because the infrared and mass singularities have been subtracted 
consistently, we should (and do) find 
$\lim_{m_0\to0}W_h^{[1]}=W_l^{[1]}$.
Numerical results for $W_h^{[1]}$ and $W_l^{[1]}$ are
in Sec.~\ref{sec:Result}.

\subsection{Mean field improvement}

Mean field improvement~\cite{LM93} has been employed extensively
in Monte Carlo work to improve tree-level estimates of couplings.
The approximation works, because most of the one-loop coefficients can 
be traced, via tadpole diagrams, to a mean field.
On the anisotropic lattice, the mean field values of the link 
variables can be defined individually for the temporal and the spatial 
links.
We denote them by $u_{\tau}$ and $u_{\sigma}$ respectively.
Then mean field improvement is achieved by replacing the link 
variables with~\cite{Kla99,TARO00,Ume00,Chen00,CPPACS00}
\begin{equation}
	U_{4} \rightarrow U_{4}/u_{\tau},  \hspace{1cm}
	U_{j} \rightarrow U_{j}/u_{\sigma}
    \hspace{0.5cm} \mbox{($j=\mbox{1--3}$)}.
\label{eq:MFreplace}
\end{equation}
With mean field improvement, the one-loop counter-terms of $u_\tau$ 
and $u_\sigma$ must be removed from perturbative coefficients.

Here we consider generically the $O(g^2)$ contributions from the 
mean field to the self-energy.
From the Feynman rules in Appendix~\ref{sec:Feynman}, the self-energy 
and its first derivative with respect to $p_4$, on the mass shell 
$({p_4},\vec{p}) = (iM_1,\veg{0})$, are
\begin{eqnarray}
	\Sigma_{\rm MF}^{[1]}(iM_1,\veg{0}) & = &
		- g^2 u^{[1]}_{\sigma} 3 r\zeta 
		- g^2 u^{[1]}_{\tau}   e^{M_1^{[1]}a_\tau} \\
	\bar{\Sigma}_{\rm MF}^{[1]}(iM_1,\veg{0}) & = &
		- g^2 u^{[1]}_{\tau} M_0a_\tau\\
	\dot{\Sigma}_{\rm MF}^{[1]}(iM_1,\veg{0}) & = &
		+ g^2 u^{[1]}_{\tau}   e^{M_1^{[1]}a_\tau} 
\end{eqnarray}
Then, the mean field contribution to the rest mass is
\begin{equation}
	M_{1\rm (MF)}^{[1]}a_\tau =
		\frac{M_0a_\tau}{1+M_0a_\tau} u^{[1]}_{\tau},
\end{equation}
and to the wave function renormalization factor
\begin{equation}
	W_{\rm MF}^{[1]} = \frac{1}{1+M_0a_\tau} u^{[1]}_{\tau},
\end{equation}
which holds for massive and massless ($M_0=0$) quarks.
The explicit values of $u_{\sigma}^{[1]}$ and $u_{\tau}^{[1]}$ depend 
on the definition of the mean field.
Since one can easily incorporate the contributions from the mean field 
improvement to the one-loop coefficients, we do not employ a specific 
scheme and quote only the contributions from the loop integrations.

\subsection{Quark bilinear operators}
\label{sec:Bilinear}

To obtain improved matrix elements, operators also must be
improved~\cite{Hea91}.
As discussed in Sec.~\ref{sec:Formulation}, with the choice $r=1/\xi$
only the multiplicative factor 
$\exp(M_1a_{\tau}/2)=1+\frac{1}{2}M_1a_{\tau}+O(a_{\tau}^2)$ 
is required at the tree-level.
In particular, with $r=1/\xi$ no new dimension-four operator is needed 
to achieve tree-level improvement.
At higher loop order the counterpart of the mass-dependent factor 
comes from the quark self-energy through the wave function factor, as 
seen in~(\ref{Z2}), and dimension-four terms are needed.

Because the tree-level rotation coefficient $d_1$ vanishes as
$m_0a_{\tau}\to 0$, we consider here currents of the form
\begin{equation}
	J_{\Gamma} (x) = \bar{\psi}_l\Gamma \psi_h (x),
\end{equation}
where $\psi_l$ and $\psi_h$ are the light and heavy quark fields 
respectively.
We consider the vector and axial vector currents, so the the $4\times4$
matrix $\Gamma$ is one of $\gamma_4$~($V_4$), $\gamma_j$~($V_j$),
$\gamma_5\gamma_4$~($A_4$), and $\gamma_5\gamma_j$~($A_j$).
We seek the matching factors $Z_{J_\Gamma}$ such that 
$Z_{J_\Gamma}J_\Gamma$ has the same matrix elements
(for $\vec{p}a_\sigma\ll 1$) as the continuum bilinear.
These matching factors are composed of two parts: the wave function of 
each quark field and the correction to the vertex.
Since the former is already obtained in previous subsection,
here we discuss the vertex corrections.

The vertex function $\Lambda_\Gamma$, which is the sum of one-particle
irreducible three-point diagrams, can be expanded
\begin{equation}
	\Lambda_\Gamma = 1 +
		\sum_{l=1}^\infty g^{2l}\Lambda_{\Gamma}^{[l]}.
\end{equation}
As with the wave function renormalization, $\Lambda_\Gamma$ is gauge
dependent and suffers from infrared and mass singularities.
For the one-loop term we again subtract of the divergent part,
\begin{equation}
	\Lambda_\Gamma^{[1]} = V_{\Gamma}^{[1]}+L_{\Gamma}^{[1]}, 
\end{equation}
where, in Feynman gauge,
\begin{equation}
	L_{\Gamma}^{[1]} = \frac{C_F}{16\pi^2}
            \left(-\half (HG-1)-\ln(\lambda^2 a_{\tau}^2)\right).
\end{equation}
The constants are again taken from the continuum expression,
so $HG=-2$ for temporal components of the currents
and $HG=2$ for spatial  components.

The sought-after matching factor $Z_{J_\Gamma}$ is simply the ratio
of the lattice and continuum radiative corrections:
\begin{equation}
	Z_{J_\Gamma} = \frac{%
	\left[Z_{2h}^{1/2}\Lambda_\Gamma Z_{2l}^{1/2}\right]^{\rm cont}}{%
	\left[Z_{2h}^{1/2}\Lambda_\Gamma Z_{2l}^{1/2}\right]^{\rm lat}}.
\end{equation}
In view of the mass dependence, we write
\begin{equation}
	(1+M_0a_\tau)^{1/2}Z_{J_\Gamma} =
		1 + \sum_{l=1}^\infty g^{2l} Z_{J_\Gamma}^{[l]},
\end{equation}
so that the $Z_{J_\Gamma}^{[l]}$ are only mildly mass dependent.
At the one-loop level we have consistently defined the finite lattice
parts so that
\begin{equation}
	Z_{J_\Gamma}^{[1]} = - \left( \half W_h^{[1]} +
		V_{\Gamma}^{[1]} + \half  W_l^{[1]} \right) 
	\label{eq:Z}
\end{equation}
is the desired one-loop coefficient of the matching factor.
It is gauge invariant and independent of the scheme for regulating the
infrared and (light-quark) mass singularities.
Numerical results for $V_\Gamma^{[1]}$ and $Z_{J_\Gamma}^{[1]}$ are in
Sec.~\ref{sec:Result}.

\section{Numerical results of one-loop perturbation theory}
\label{sec:Result}

In this section we present our results for the one-loop coefficients.
They are obtained numerically using the Monte Carlo integration
program {\tt BASES}~\cite{Kaw95}.
We give one-loop terms
for the rest mass, {\it i.e.}, $M_1^{[1]}$ and $c^{[1]}$;
for the infrared-finite parts of the wave function renormalization
factors and the vertex functions, {\it i.e.},
$W_l^{[1]}$, $W_h^{[1]}$ and $V_\Gamma^{[1]}$;
and for the currents' matching factors~$Z_{J_\Gamma}^{[1]}$.
In this section we are concerned with zero three-momentum, so for
brevity we set the temporal lattice spacing $a_\tau=1$.
When the spatial lattice spacing is needed, we use the
anisotropy~$\xi$.

The spatial and the temporal parts of $M_1^{[1]}$ are 
listed separately, namely
\begin{equation}
	g^2 M_1^{[1]} \to
		g^2_{\tau}M^{[1]}_{1\tau} + g^2_{\sigma}M^{[1]}_{1\sigma},
\end{equation}
so that one could use different (improved) couplings in a practical 
evaluation of the perturbative rest mass.
On the other hand, for the other quantities we show the combined values
of the spatial and the temporal parts, because we are interested mostly
in seeing how they behave when $M_0\xi\sim1$ while
$M_0:=M_0a_\tau$~small.

\begin{figure}[bp]
\centerline{
\leavevmode\psfig{file=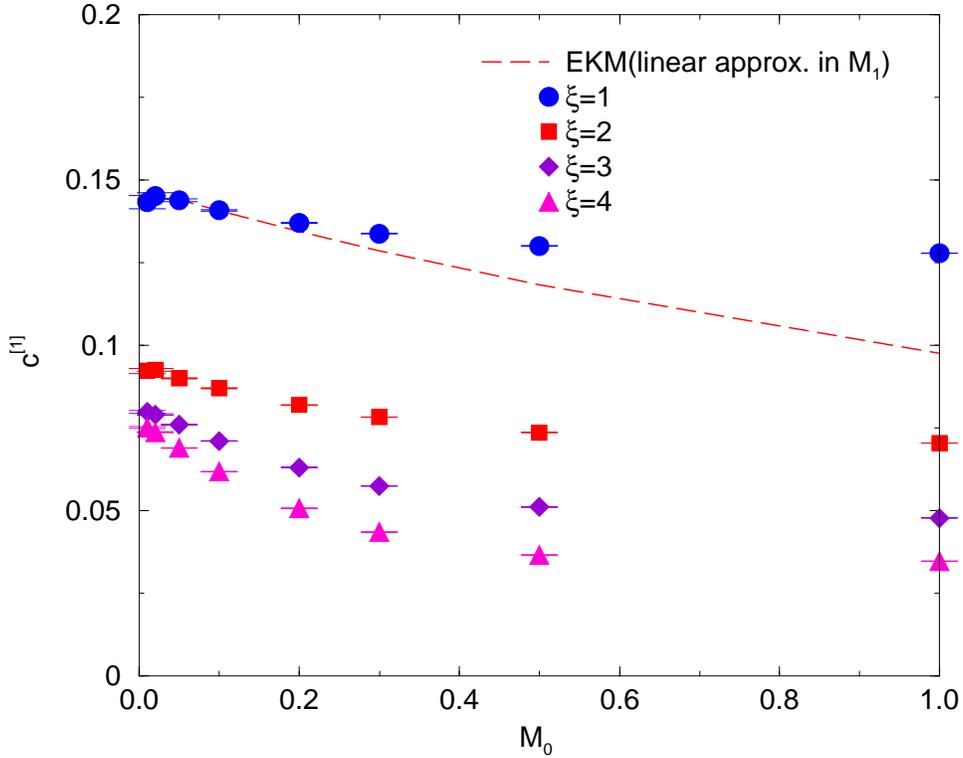,width=\figwidth} }
\caption{The non-logarithmic part of the mass renormalization
factor~$c^{[1]}$.
The dashed line is the linear approximation, based on
Ref.~\cite{MKE98}.}
\label{fig:z_m}
\end{figure}

The spatial and the temporal parts of $M_1^{[1]}$, $M_{1\sigma}^{[1]}$
and $M_{1\tau}^{[1]}$ respectively, are listed in Table~\ref{tab:mass}
for a range of~$M_0<1$ at four values of~$\xi$: 1, 2, 3 and~4.
\begin{table}[bp]
\begin{center}
\caption{The one-loop correction to the rest mass, $M_1^{[0]}$,
for various values of $M_0$ at four values of $\xi$.}
\label{tab:mass}
\vspace{3mm}
\begin{tabular}{clllll} 
\hline\hline
& $M_0$ &$\xi=1$ & $\xi=2$ & $\xi=3$ & $\xi=4$ \\
\hline
$M_{1\sigma}^{[1]}$ 
 & 0.01 & 0.00111(3) & 0.00151(1)  & 0.001477(7) & 0.001456(5) \\
 & 0.02 & 0.00286(3) & 0.00266(1)  & 0.002571(7) & 0.002534(5) \\
 & 0.05 & 0.00624(3) & 0.00543(1)  & 0.005116(7) & 0.004958(5) \\
 & 0.10 & 0.01036(3) & 0.00868(1)  & 0.008017(7) & 0.007563(5) \\
 & 0.20 & 0.01620(3) & 0.01282(1)  & 0.011177(7) & 0.009938(5) \\
 & 0.30 & 0.02002(3) & 0.01504(1)  & 0.012449(6) & 0.010507(4) \\
 & 0.50 & 0.02434(2) & 0.01665(1)  & 0.012582(6) & 0.009810(4) \\
 & 1.00 & 0.02694(2) & 0.015300(9) & 0.009960(4) & 0.006959(3) \\
\hline
$M_{1\tau}^{[1]}$ 
 & 0.01 & 0.00251(1) & 0.002018(5) & 0.001895(3) & 0.001847(2) \\
 & 0.02 & 0.00469(1) & 0.003674(6) & 0.003423(3) & 0.003333(3) \\
 & 0.05 & 0.01049(2) & 0.007864(7) & 0.007236(4) & 0.006959(3) \\
 & 0.10 & 0.01878(2) & 0.013606(7) & 0.012275(5) & 0.011562(4) \\
 & 0.20 & 0.03222(2) & 0.022430(8) & 0.019480(5) & 0.017757(4) \\
 & 0.30 & 0.04305(2) & 0.029095(7) & 0.024581(6) & 0.021753(5) \\
 & 0.50 & 0.05988(2) & 0.038651(9) & 0.031177(6) & 0.026528(6) \\
 & 1.00 & 0.08646(2) & 0.05215(1)  & 0.039395(7) & 0.031916(6) \\
\hline\hline
\end{tabular}
\end{center}
\end{table}
We plot $c^{[1]}$ vs.~$M_0$ in Fig.~\ref{fig:z_m},
and the numerical values are given in Table~\ref{tab:z_m}.
\begin{table}[tp]
\begin{center}
\caption{The non-logarithmic part of the mass renormalization
factor~$c^{[1]}$.}
\label{tab:z_m}
\vspace{3mm}
\begin{tabular}{clllll} 
\hline\hline
	& $M_0$ &$\xi=1$ & $\xi=2$ & $\xi=3$ & $\xi=4$ \\
\hline
$c^{[1]}$ 
 & 0.01 & 0.143(2)   & 0.092(1)   & 0.0798(4)  & 0.0751(3) \\
 & 0.02 & 0.145(1)   & 0.092(1)   & 0.0790(2)  & 0.0738(2) \\
 & 0.05 & 0.1438(4)  & 0.0900(2)  & 0.0760(1)  & 0.0690(1) \\
 & 0.10 & 0.1408(2)  & 0.08701(8) & 0.07104(6) & 0.06181(4) \\
 & 0.20 & 0.1370(1)  & 0.08197(5) & 0.06309(3) & 0.05070(2) \\
 & 0.30 & 0.13374(7) & 0.07828(4) & 0.05744(2) & 0.04354(2) \\
 & 0.50 & 0.13004(5) & 0.07365(2) & 0.05110(1) & 0.03662(1) \\
 & 1.00 & 0.12787(3) & 0.07043(1) & 0.04781(1) & 0.03473(1) \\
\hline\hline
\end{tabular}
\end{center}
\end{table}
One sees that the mass dependence is significant, but not drastic.

\begin{figure}[bp]
\centerline{
\leavevmode\psfig{file=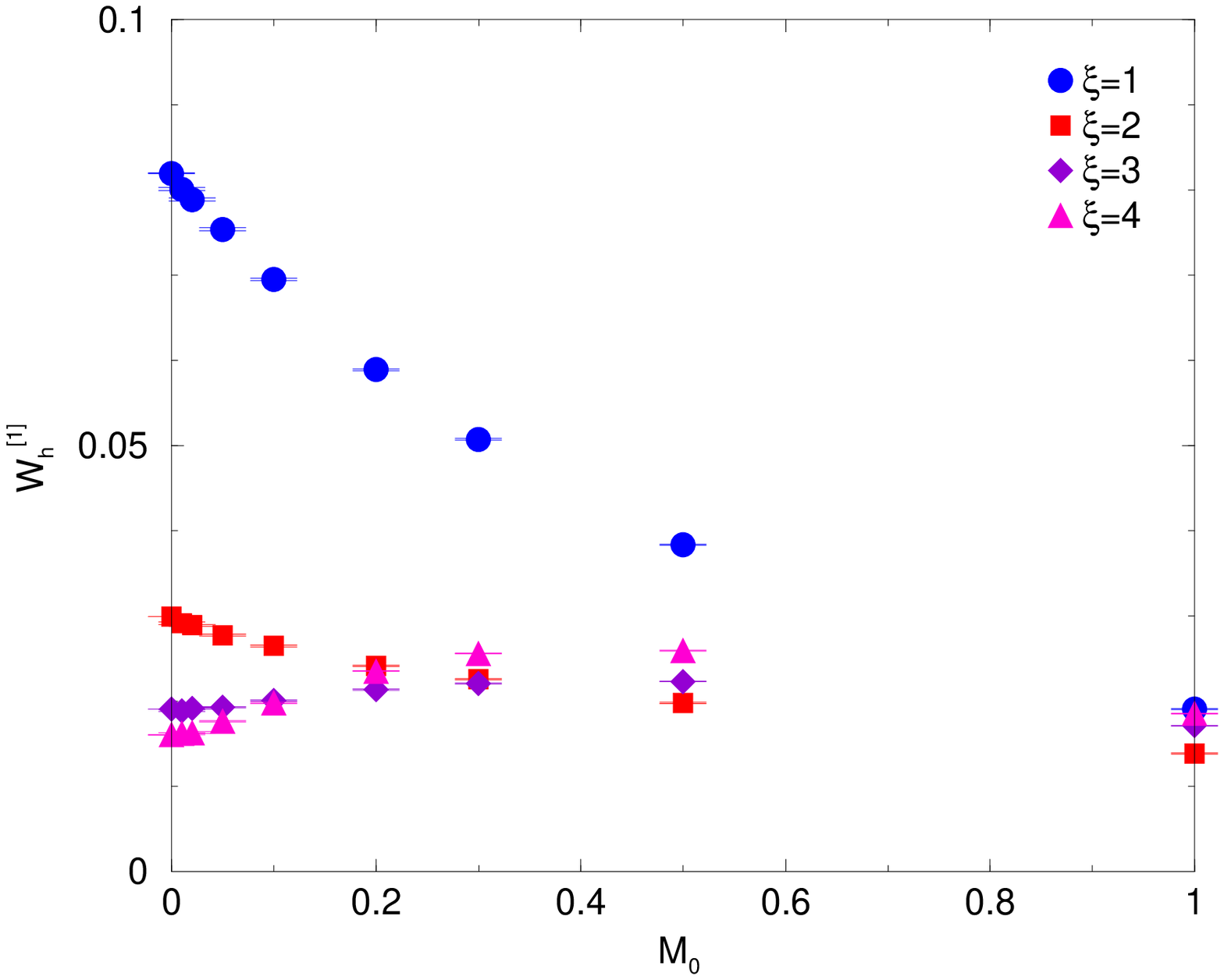,width=\figwidth} }
\caption{Mass dependence and $\xi$ dependence of the one-loop correction 
to the heavy quark wave function.}
\label{fig:Wh}
\end{figure}

Table~\ref{tab:wave} lists the one-loop corrections~$W_l^{[1]}$ 
\begin{table}[tp]
\begin{center}
\caption{The one-loop correction to the light and the heavy
quark wave functions.}
\label{tab:wave}
\vspace{3mm}
\begin{tabular}{clllll} 
\hline\hline
& $M_0$ & $\xi=1$ & $\xi=2$ & $\xi=3$ & $\xi=4$ \\
\hline
$W_l^{[1]}$
& 0.00 & 0.08194(2)   &  0.02994(1) & 0.01911(1) &  0.01605(1)\\
\hline
$W_h^{[1]}$
& 0.01 & 0.08009(19) & 0.02913(13) & 0.01890(11) & 0.01621(11) \\
& 0.02 & 0.07887(18) & 0.02892(12) & 0.01918(10) & 0.01634(9) \\
& 0.05 & 0.07537(15) & 0.02774(11) & 0.01929(8)  & 0.01765(7) \\
& 0.10 & 0.06949(14) & 0.02649(9)  & 0.02009(7)  & 0.01981(6) \\
& 0.20 & 0.05892(11) & 0.02417(7)  & 0.02137(5)  & 0.02354(5) \\
& 0.30 & 0.05072(11) & 0.02258(6)  & 0.02209(5)  & 0.02561(4) \\
& 0.50 & 0.03833(7)  & 0.01979(5)  & 0.02233(4)  & 0.02594(4) \\
& 1.00 & 0.01908(7)  & 0.01388(5)  & 0.01715(5)  & 0.01856(5) \\
\hline\hline
\end{tabular}
\end{center}
\end{table}
and~$W_h^{[1]}$ to the massless and massive quark wave function 
renormalization factors.
The mass dependence is shown in Fig.~\ref{fig:Wh}.
Here the introduction of anisotropy is seen to reduce the mass
dependence greatly.
Since $W_h^{[1]}$ connects smoothly to $W_l^{[1]}$, one sees that we 
have subtracted the infrared singularities in a consistent~way.

\begin{figure}[bp]
\centerline{
\leavevmode\psfig{file=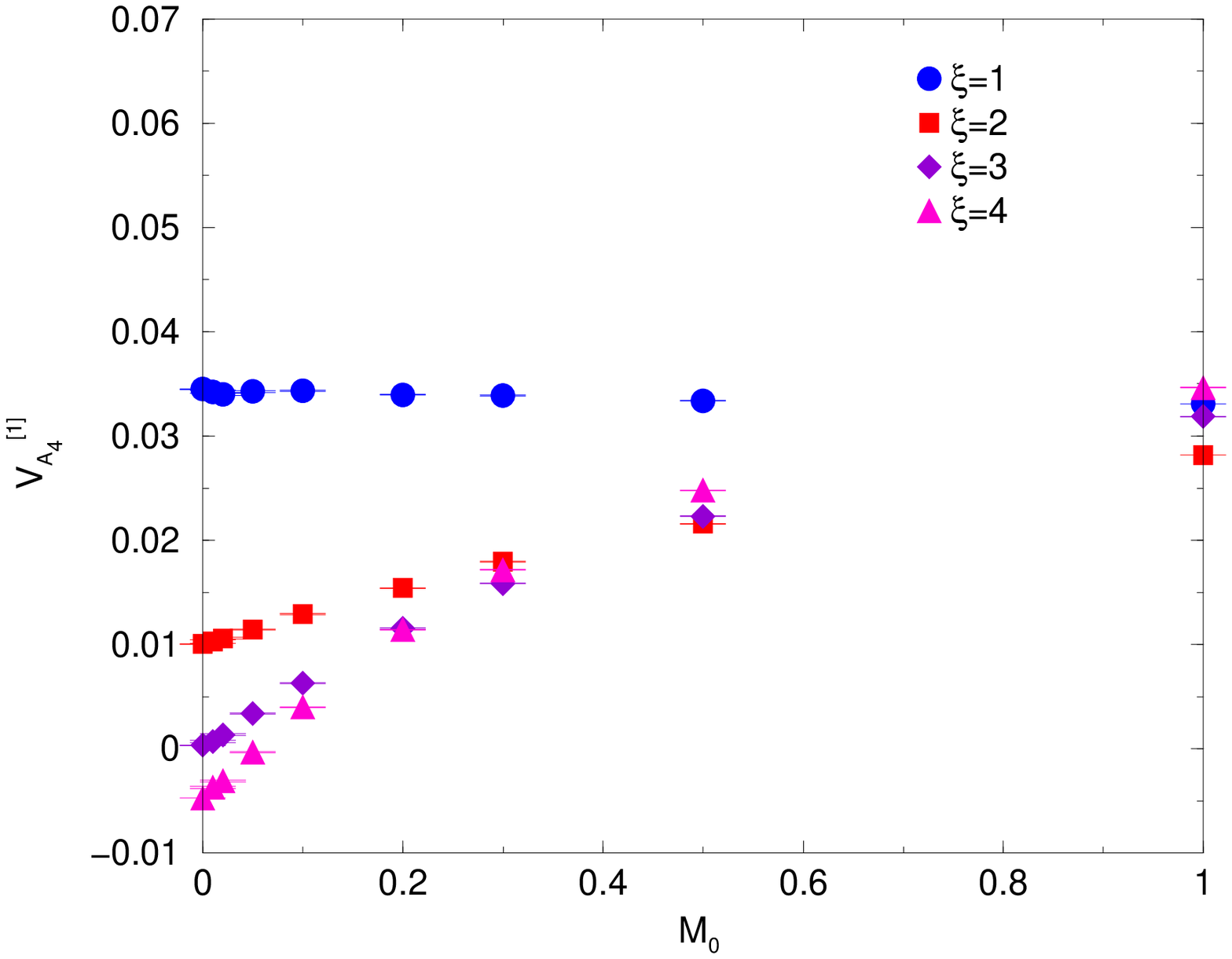,width=\figwidth}} 
\centerline{
\leavevmode\psfig{file=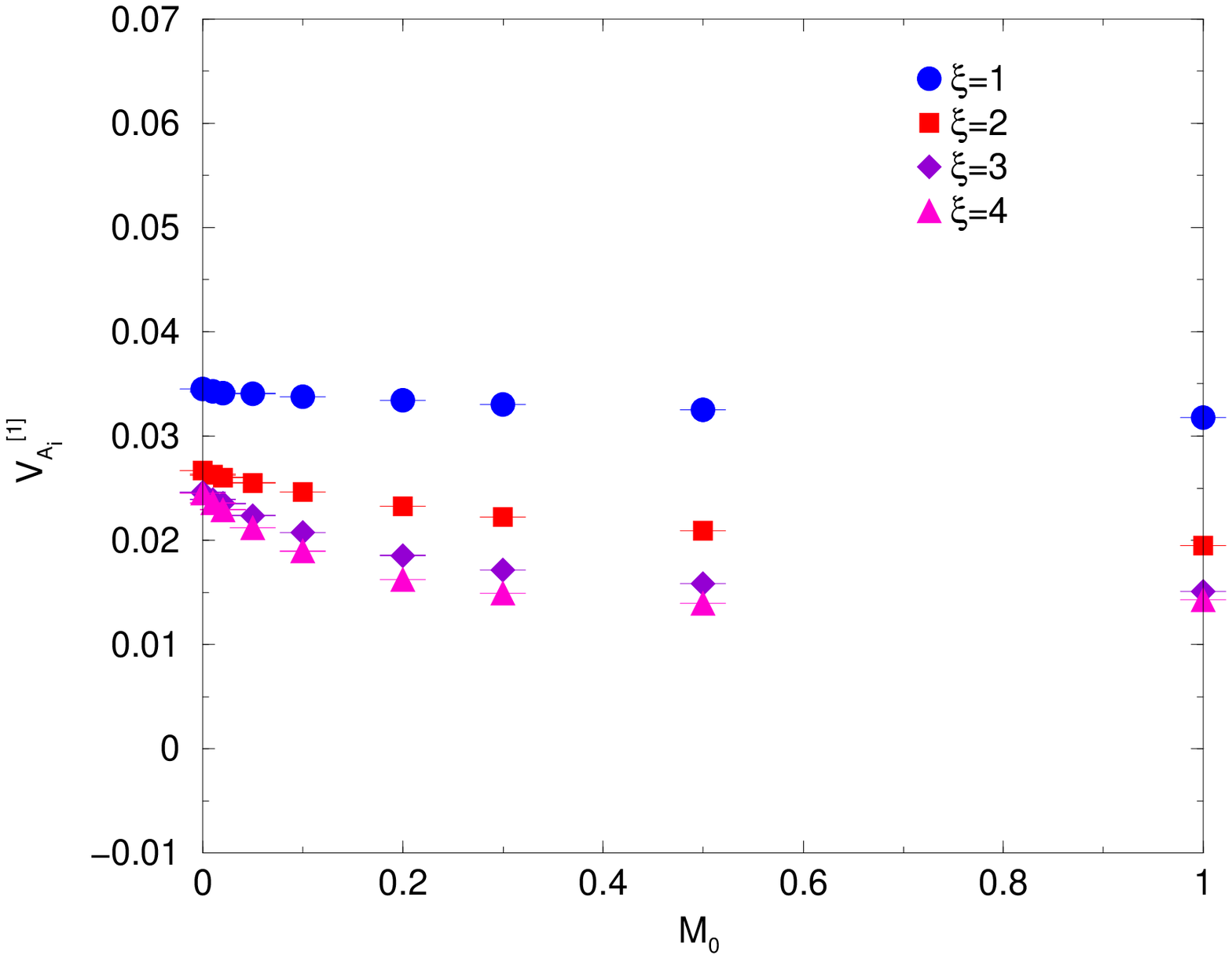,width=\figwidth}} 
\caption{Mass dependence and $\xi$ dependence of the one-loop 
vertex corrections to the axial vector current.}
\label{fig:V4}
\end{figure}

\begin{figure}[hp]
\centerline{
\leavevmode\psfig{file=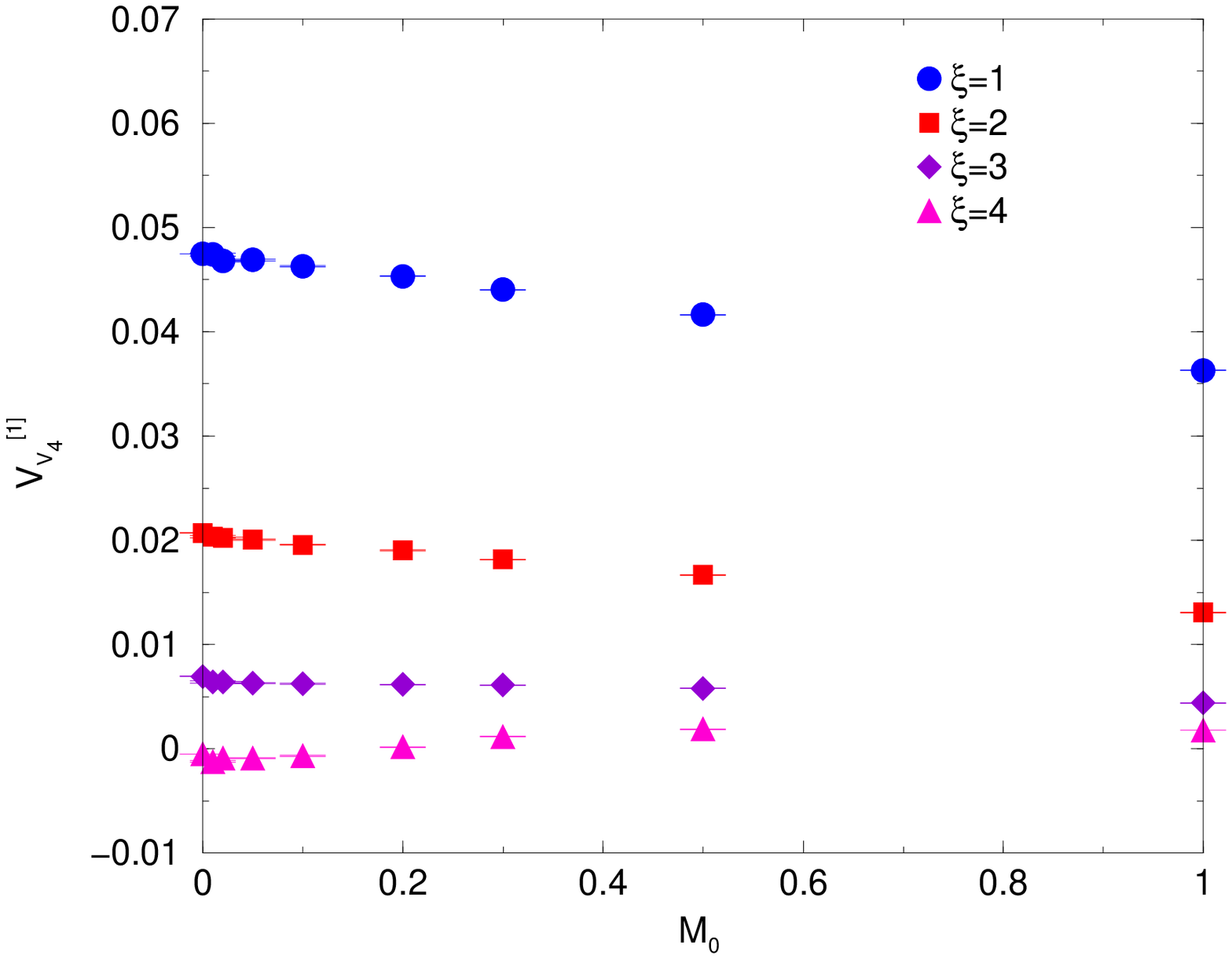,width=\figwidth}} 
\centerline{
\leavevmode\psfig{file=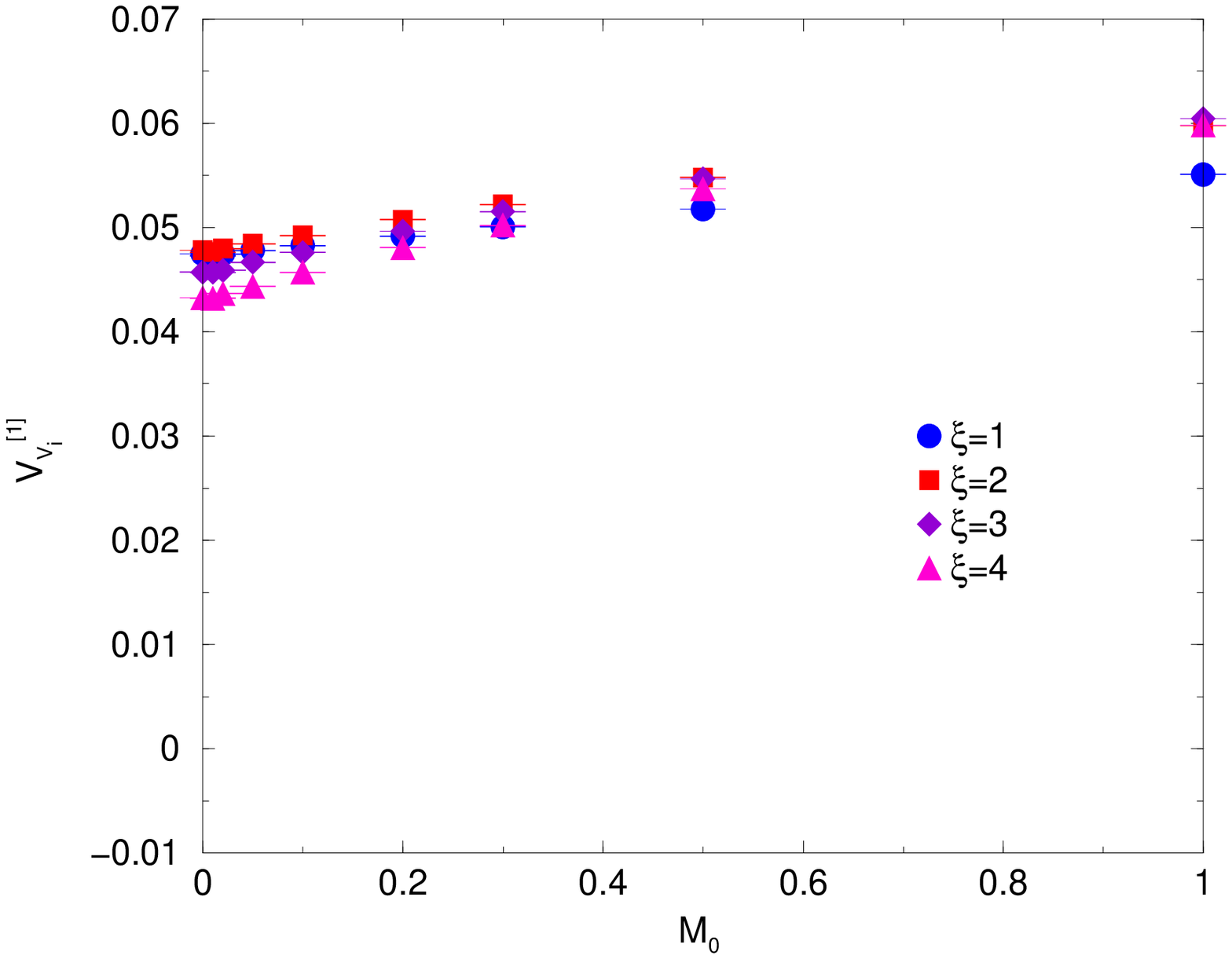,width=\figwidth}} 
\caption{Mass dependence and $\xi$ dependence of the one-loop 
vertex corrections to the vector current.}
\label{fig:VV}
\end{figure}

Tables~\ref{tab:va} and~\ref{tab:vv} list the one-loop corrections to
the axial vector and the vector current vertex functions, respectively.
\begin{table}[tp]
\begin{center}
\caption{The one-loop correction to the axial vector current vertex 
function.}
\label{tab:va}
\vspace{3mm}
\begin{tabular}{clllll} 
\hline\hline
 & $M_0$ & $\xi=1$ & $\xi=2$ & $\xi=3$ & $\xi=4$ \\
\hline
$V_{A_4}^{[1]}$
 & 0.00 & 0.03449(1)  & 0.01005(1)  & 0.00033(1)  & $-$0.00472(1)  \\
 & 0.01 & 0.03425(13) & 0.01028(17) & 0.00071(10) & $-$0.00369(10) \\
 & 0.02 & 0.03399(10) & 0.01061(10) & 0.00134(8)  & $-$0.00309(8)  \\
 & 0.05 & 0.03426(8)  & 0.01145(7)  & 0.00337(6)  & $-$0.00036(6)  \\
 & 0.10 & 0.03435(6)  & 0.01294(5)  & 0.00628(5)  & 0.00399(5)  \\
 & 0.20 & 0.03396(4)  & 0.01294(4)  & 0.01155(3)  & 0.01142(3)  \\
 & 0.30 & 0.03389(4)  & 0.01543(4)  & 0.01585(3)  & 0.01717(3)  \\
 & 0.50 & 0.03337(3)  & 0.02160(2)  & 0.02230(2)  & 0.02479(2)  \\
 & 1.00 & 0.03309(2)  & 0.02818(2)  & 0.03187(2)  & 0.03465(1)  \\
\hline
$V_{A_i}^{[1]}$
 & 0.00 & 0.03450(1) & 0.02669(1) & 0.02460(1) & 0.02454(1) \\
 & 0.01 & 0.03428(5) & 0.02629(4) & 0.02391(3) & 0.02359(3) \\
 & 0.02 & 0.03410(3) & 0.02602(3) & 0.02352(2) & 0.02294(2) \\
 & 0.05 & 0.03407(2) & 0.02551(2) & 0.02237(2) & 0.02121(2) \\
 & 0.10 & 0.03378(2) & 0.02463(2) & 0.02075(1) & 0.01896(1) \\
 & 0.20 & 0.03342(1) & 0.02326(1) & 0.01854(1) & 0.01624(1) \\
 & 0.30 & 0.03302(1) & 0.02223(1) & 0.01716(1) & 0.01490(1) \\
 & 0.50 & 0.03253(1) & 0.02093(1) & 0.01582(1) & 0.01397(1) \\
 & 1.00 & 0.03176(1) & 0.01950(1) & 0.01512(1) & 0.01429(1) \\
\hline\hline
\end{tabular}
\end{center}
\end{table}
\begin{table}[tp]
\begin{center}
\caption{The one-loop correction to the vector current vertex 
function.}
\label{tab:vv}
\vspace{3mm}
\begin{tabular}{clllll} 
\hline\hline
& $M_0$ & $\xi=1$ & $\xi=2$ & $\xi=3$ & $\xi=4$ \\
\hline
$V_{V_4}^{[1]}$
 & 0.00 & 0.04749(1)  & 0.02071(1)  & 0.00695(1)  & $-$0.00049(1)  \\
 & 0.01 & 0.04740(13) & 0.02036(11) & 0.00641(10) & $-$0.00122(10)  \\
 & 0.02 & 0.04677(10) & 0.02021(9)  & 0.00641(9)  & $-$0.00090(8)  \\
 & 0.05 & 0.04690(8)  & 0.02006(6)  & 0.00631(6)  & $-$0.00090(6)  \\
 & 0.10 & 0.04628(7)  & 0.01957(6)  & 0.00622(5)  & $-$0.00065(7)  \\
 & 0.20 & 0.04532(5)  & 0.01904(4)  & 0.00617(3)  & $-$0.00016(3)  \\
 & 0.30 & 0.04404(4)  & 0.01815(3)  & 0.00611(3)  & $-$0.00116(3)  \\
 & 0.50 & 0.04165(3)  & 0.01669(3)  & 0.00579(2)  & $-$0.00188(2)  \\
 & 1.00 & 0.03629(3)  & 0.01308(2)  & 0.00439(2)  & $-$0.00179(1)  \\
\hline
$V_{V_i}^{[1]}$
 & 0.00 & 0.04748(1) & 0.04784(1) & 0.04573(1) & 0.04326(1) \\
 & 0.01 & 0.04727(4) & 0.04764(4) & 0.04570(3) & 0.04320(3) \\
 & 0.02 & 0.04746(3) & 0.04797(3) & 0.04589(3) & 0.04366(2) \\
 & 0.05 & 0.04779(2) & 0.04843(2) & 0.04665(2) & 0.04436(2) \\
 & 0.10 & 0.04825(2) & 0.04923(2) & 0.04762(1) & 0.04568(1) \\
 & 0.20 & 0.04917(1) & 0.05075(1) & 0.04967(1) & 0.04807(1) \\
 & 0.30 & 0.05007(1) & 0.05219(1) & 0.05151(1) & 0.05020(1) \\
 & 0.50 & 0.05175(1) & 0.05481(1) & 0.05467(1) & 0.05369(1) \\
 & 1.00 & 0.05510(1) & 0.05977(1) & 0.06046(1) & 0.05980(1) \\
\hline\hline
\end{tabular}
\end{center}
\end{table}
The mass dependence is shown in Fig.~\ref{fig:V4} and~\ref{fig:VV}.
For the axial vector current, the mass dependence with anisotropy is
larger than with $\xi=1$, but still small.

\begin{figure}[bp]
\centerline{
\leavevmode\psfig{file=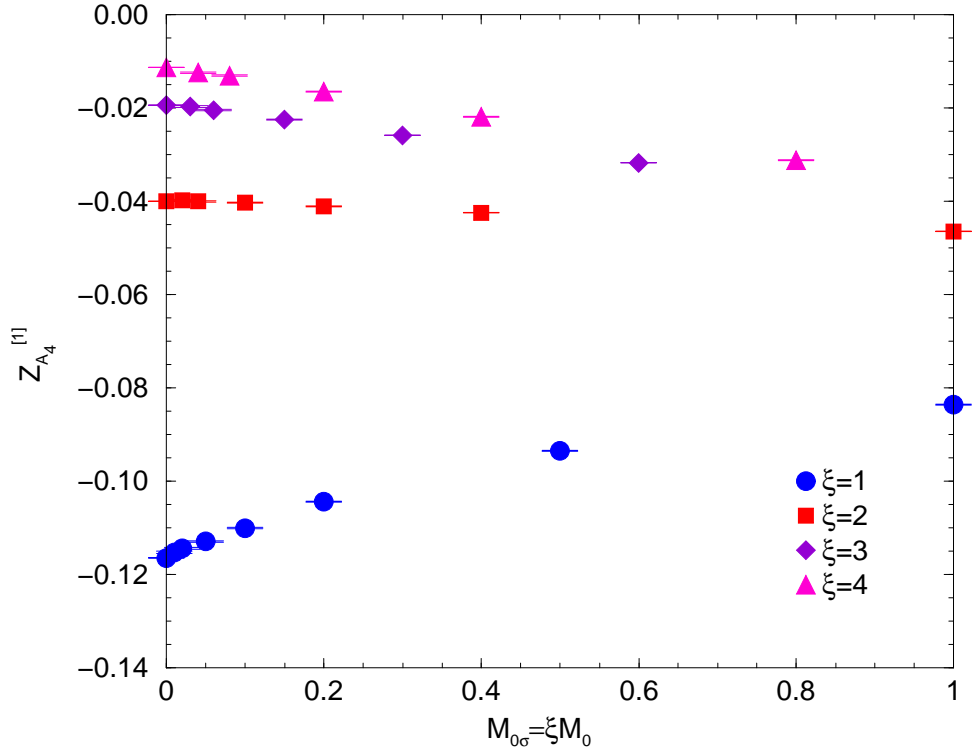,width=\figwidth} }
\centerline{
\leavevmode\psfig{file=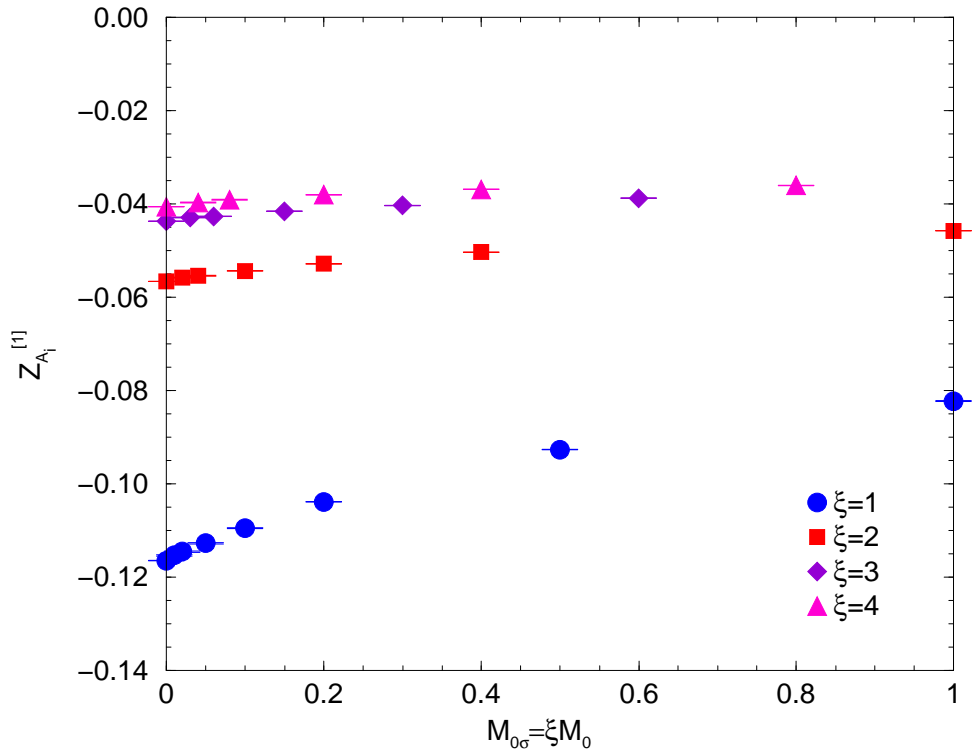,width=\figwidth} }
\caption{
$Z_{A_4}^{[1]}$ and $Z_{A_i}^{[1]}$ vs.\ $M_{0\sigma}$.}
\label{fig:za}
\end{figure}

\begin{figure}[bp]
\centerline{
\leavevmode\psfig{file=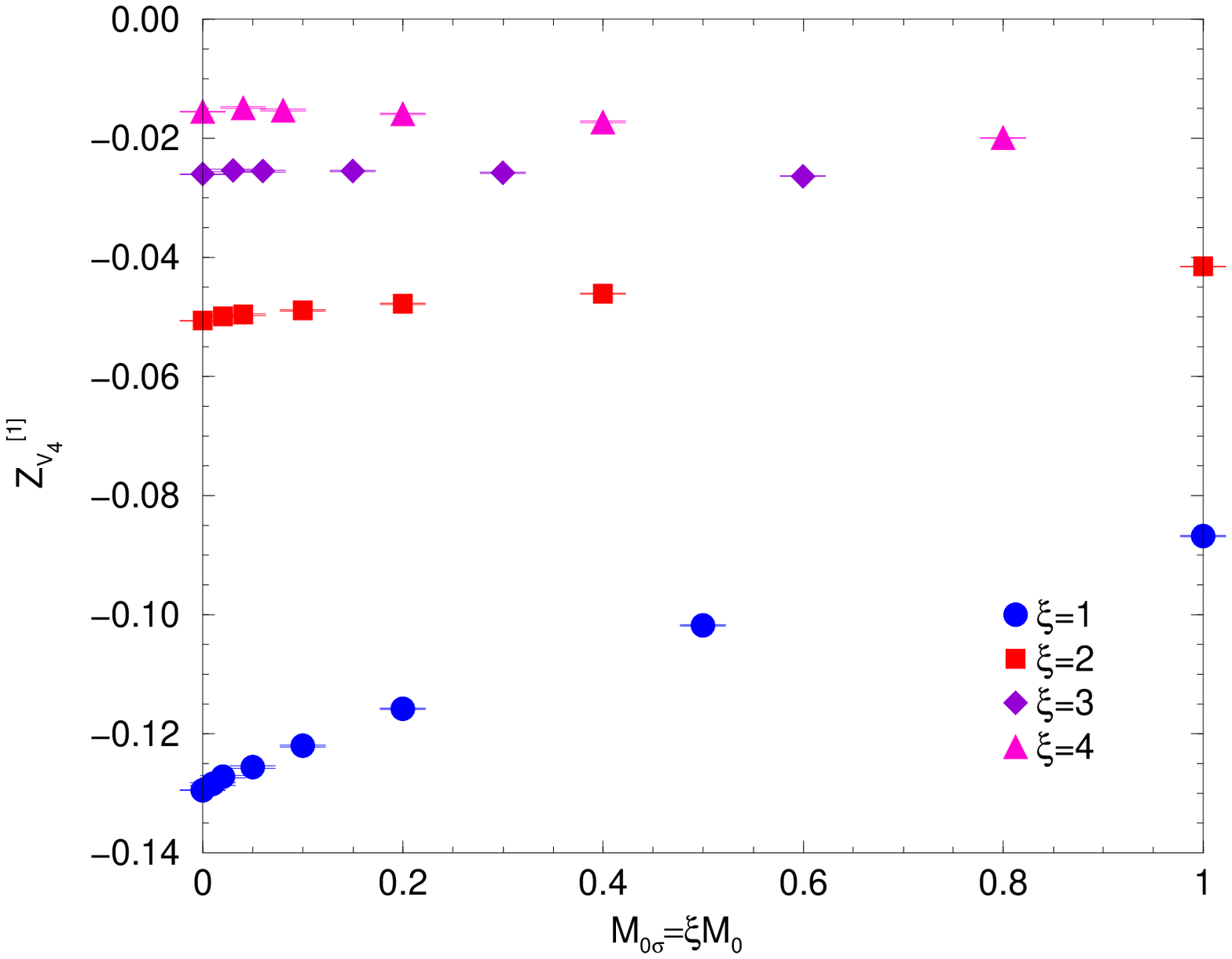,width=\figwidth} }
\centerline{
\leavevmode\psfig{file=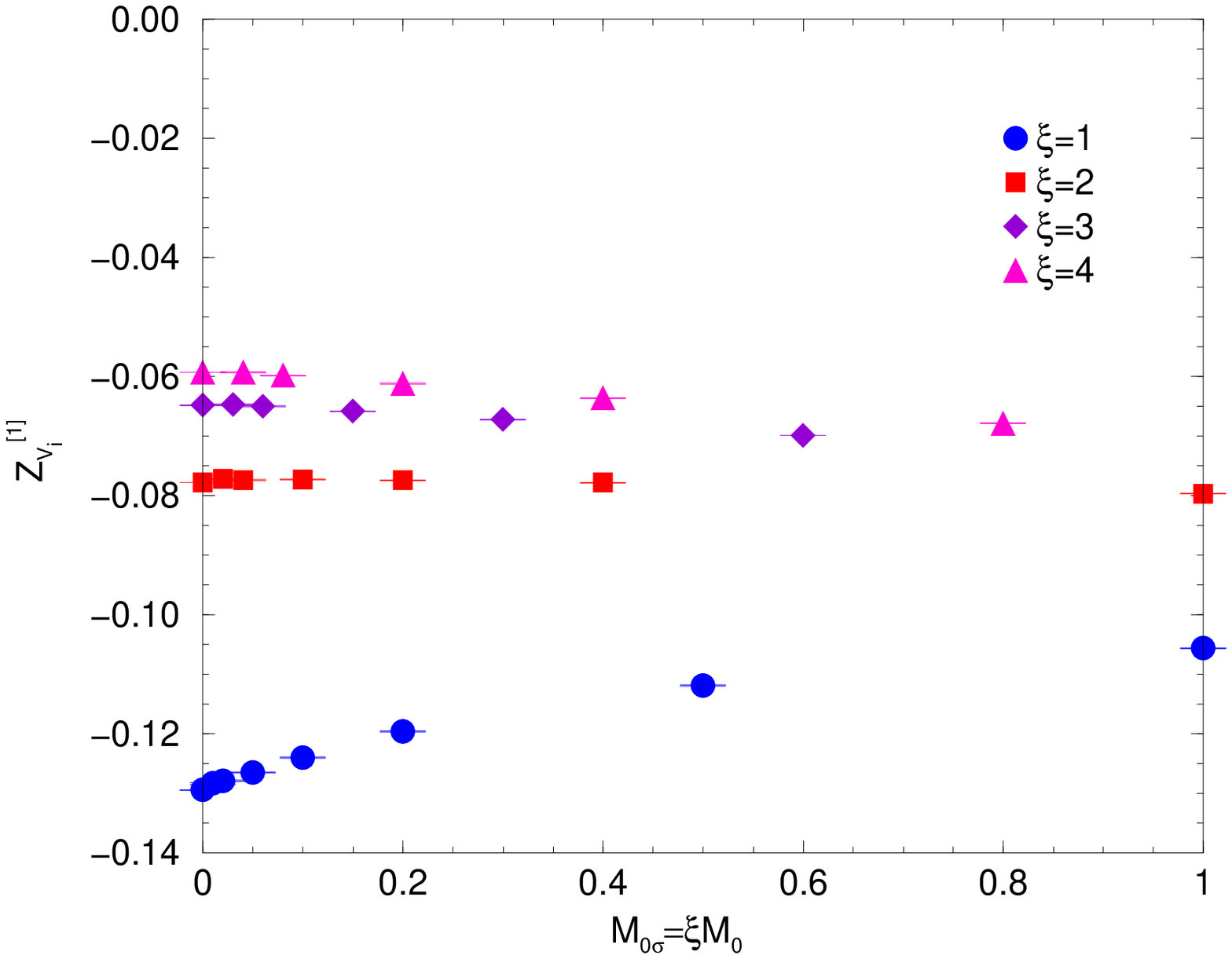,width=\figwidth} }
\caption{
$Z_{V_4}^{[1]}$ and $Z_{V_i}^{[1]}$ vs.\ $M_{0\sigma}$.}
\label{fig:zv}
\end{figure}

These results are combined to obtain the one-loop part of
the matching factors according to Eq.~(\ref{eq:Z}).
The results are listed in Table~\ref{tab:Z} and plotted in
Figs.~\ref{fig:za} and~\ref{fig:zv}.
\begin{table}[p]
\begin{center}
\caption{The one-loop term of the matching factor between the
continuum and the lattice theories for the axial vector
and the vector currents.
The spatial and the temporal contributions are summed.}
\label{tab:Z}
\vspace{3mm}
\begin{tabular}{clllll}
\hline\hline
 & $M_0$ & $\xi=1$ & $\xi=2$ & $\xi=3$ & $\xi=4$ \\
\hline
$Z_{A_4}^{[1]}$
 & 0.00 &$-$0.11643(3)&$-$0.04000(3)&$-$0.01944(3)&$-$0.01133(2) \\
 & 0.01 &$-$0.1153(2) &$-$0.0398(2) &$-$0.0197(2) &$-$0.0124(2)  \\
 & 0.02 &$-$0.1144(2) &$-$0.0400(2) &$-$0.0205(1) &$-$0.0131(1)  \\
 & 0.05 &$-$0.1129(2) &$-$0.0403(1) &$-$0.0226(1) &$-$0.0165(1)  \\
 & 0.10 &$-$0.1101(1) &$-$0.0412(1) &$-$0.02588(8)&$-$0.02192(9) \\
 & 0.20 &$-$0.1044(1) &$-$0.04248(8)&$-$0.03178(6)&$-$0.03122(6) \\
 & 0.30 &$-$0.1002(1) &$-$0.04423(8)&$-$0.03644(6)&$-$0.03800(5) \\
 & 0.50 &$-$0.09351(8)&$-$0.04646(6)&$-$0.04302(5)&$-$0.04579(5) \\
 & 1.00 &$-$0.08360(7)&$-$0.05009(5)&$-$0.04999(4)&$-$0.05196(4) \\
\hline
$Z_{A_i}^{[1]}$
 & 0.00 &$-$0.11644(3)&$-$0.05664(2)&$-$0.04371(2)&$-$0.04059(1) \\
 & 0.01 &$-$0.1153(2) &$-$0.0558(1) &$-$0.04291(9)&$-$0.03972(9) \\
 & 0.02 &$-$0.1145(1) &$-$0.05545(9)&$-$0.04267(8)&$-$0.03914(7) \\
 & 0.05 &$-$0.1127(1) &$-$0.05436(8)&$-$0.04157(6)&$-$0.03807(6) \\
 & 0.10 &$-$0.1095(1) &$-$0.05285(7)&$-$0.04035(5)&$-$0.03689(5) \\
 & 0.20 &$-$0.10385(8)&$-$0.05032(5)&$-$0.03877(4)&$-$0.03604(4) \\
 & 0.30 &$-$0.09935(7)&$-$0.04849(5)&$-$0.03776(4)&$-$0.03573(4) \\
 & 0.50 &$-$0.09266(6)&$-$0.04580(4)&$-$0.03654(4)&$-$0.03497(3) \\
 & 1.00 &$-$0.08227(5)&$-$0.04142(4)&$-$0.03324(4)&$-$0.03160(3) \\
\hline
$Z_{V_4}^{[1]}$
 & 0.00 &$-$0.12943(4)&$-$0.05066(2)&$-$0.02606(2)&$-$0.01556(3) \\
 & 0.01 &$-$0.1284(2) &$-$0.0499(2) &$-$0.0254(2) &$-$0.0149(2)  \\
 & 0.02 &$-$0.1272(2) &$-$0.0496(2) &$-$0.0256(1) &$-$0.0153(1)  \\
 & 0.05 &$-$0.1256(2) &$-$0.0489(1) &$-$0.0255(1) &$-$0.0160(1)  \\
 & 0.10 &$-$0.1220(2) &$-$0.0478(1) &$-$0.02582(8)&$-$0.0173(1)  \\
 & 0.20 &$-$0.1158(1) &$-$0.04610(8)&$-$0.02640(7)&$-$0.01996(6) \\
 & 0.30 &$-$0.1104(1) &$-$0.04442(7)&$-$0.02670(6)&$-$0.02199(6) \\
 & 0.50 &$-$0.10178(8)&$-$0.04156(6)&$-$0.02651(5)&$-$0.02288(5) \\
 & 1.00 &$-$0.08680(7)&$-$0.03499(5)&$-$0.02252(5)&$-$0.01910(4) \\
\hline
$Z_{V_i}^{[1]}$
 & 0.00 &$-$0.12942(3)&$-$0.07779(2)&$-$0.06484(2)&$-$0.05931(2) \\
 & 0.01 &$-$0.1283(1) &$-$0.0772(1) &$-$0.06470(9)&$-$0.05933(9) \\
 & 0.02 &$-$0.1279(1) &$-$0.07740(9)&$-$0.06503(8)&$-$0.05985(8) \\
 & 0.05 &$-$0.1264(1) &$-$0.07728(8)&$-$0.06585(7)&$-$0.06121(6) \\
 & 0.10 &$-$0.1240(1) &$-$0.07744(7)&$-$0.06721(5)&$-$0.06361(5) \\
 & 0.20 &$-$0.11960(8)&$-$0.07781(5)&$-$0.06990(4)&$-$0.06787(4) \\
 & 0.30 &$-$0.11639(8)&$-$0.07846(5)&$-$0.07211(4)&$-$0.07103(4) \\
 & 0.50 &$-$0.11188(6)&$-$0.07968(4)&$-$0.07538(4)&$-$0.07469(3) \\
 & 1.00 &$-$0.10561(5)&$-$0.08168(4)&$-$0.07859(4)&$-$0.07711(4) \\
\hline\hline
\end{tabular}
\end{center}
\end{table}
The magnitude of the one-loop correction decreases as 
$\xi$ increases, in all cases.
Figures~\ref{fig:za} and~\ref{fig:zv} are the most important results of
this section.
They are relevant to phenomenological applications to charm physics.
Moreover, these results test, at the one-loop level, whether the
matching factors are well-behaved in the interesting region with
small~$M_0$ but $M_0\xi\sim1$.
For this reason we have plotted them, as in Fig.~\ref{fig:zeta_tree},
not against~$M_0$ but~$M_0\xi$.
By inspection of Figs.~\ref{fig:za} and~\ref{fig:zv}, one can see that
the small $M_0$ Taylor series continues to be a good approximation to
the full mass dependence for all $M_0\xi\leq1$, for $\xi\geq2$.
Had we found a stronger mass dependence (like that of~$\zeta^{[0]}$ for
$r=1$), one would begin to doubt the feasibility of the ideas laid out
in Ref.~\cite{Kla99} also for our choice~$r=1/\xi$.

For $\xi=1$, our results should reproduce previous calculations
on isotropic lattices.
In the case of the mass and the wave function renormalization,
we (independently) reproduced the full mass dependence of
Ref.~\cite{MKE98}.
For the matching factors of the currents, only the result for the
massless quark is (independently) available~\cite{Aoki98}, and we find
agreement.

In conclusion, in the whole region of $M_0$ and $\xi$ we surveyed, we
found good behavior connecting the continuum limit with the region
of practical interest.
For charmed hadrons, the target region of the heavy quark mass $M_0$
is around 0.1--0.3 on lattices with anisotropy $\xi$=3--4.
The required one-loop coefficients of the renormalization factors are
easily obtained by interpolating the values in the tables using, for
example, spline interpolation.

\section{Conclusion}
\label{sec:Conclusion}

In this paper, we have studied the $O(a)$ improvement of Wilson quarks on
anisotropic lattices.
At the tree-level we find that a certain choice of the parameters,
$r=1/\xi$~\cite{TARO00,Ume00}, is well-behaved in the region of
practical interest for charmed hadrons, namely
$M_0a_\sigma\sim1$, while $M_0a_\tau$ is small.
On the other hand, with a different choice,
$r=1$~\cite{Kla99,Chen00,CPPACS00},
continuum behavior is reached only for $M_0a_\sigma\ll1$.
With this latter choice a non-relativistic
interpretation~\cite{EKM97,Kronfeld:2000ck} is still possible,
but a mass-independent renormalization, which was proposed in
Ref.~\cite{Kla99}, is obstructed.

The choice $r=1/\xi$ also simplifies tree-level $O(a)$ improvement.
The action does not require separate temporal and spatial
hopping parameters.
The currents require mass-dependent matching factors, but no
intrinsically dimension-four terms.

We therefore have started to examine the behavior of this choice at the
one-loop level.
We have computed the one-loop contributions to the rest mass and to the
matching factors of the vector and axial vector currents.
The matching factors depend significantly on~$\xi$.
A more critical observation is that they are well approximated by
Taylor expansions
\begin{equation}
	Z_{J_\Gamma}(\xi,M_0a_\tau) \simeq Z_{J_\Gamma}(\xi,0) +
		M_0a_\tau Z'_{J_\Gamma}(\xi,0)
\end{equation}
for $M_0a_\sigma=\xi M_0a_\tau\leq 1$ and $\xi=2$--$4$.
This region encompasses the one suitable for the charmed quark with
currently available computer resources.

There are several issues that remain to be studied.
The first is to compute the one-loop corrections to the ratio of
hopping parameters~$\zeta$, the clover coefficients~$c_B$ and $c_E$,
and dimension-four terms in the currents.
The calculation of~$\zeta$ is especially difficult, because it requires
the one-loop kinetic mass~$M_2^{[1]}$.
As at the tree-level, it is crucial to compute the full mass dependence,
so one can check whether low-order Taylor expansions work well for
$M_0a_\sigma\sim1$.
Only with the full mass dependence can one check whether $\xi$, which
comes with the couplings in the action, and $M_0a_\tau$, which also
comes from the on-shell condition, come together to form
$M_0a_\sigma=\xi M_0a_\tau$.
If not, then one could proceed with a non-perturbative calculation of
$\zeta(\xi,0)$, $\zeta'(\xi,0)$, $c_B(\xi,0)$, $c_E(\xi,0)$, etc.

A more practical problem is to define renormalized couplings.
The scale-setting scheme of Brodsky, Lepage, and Mackenzie (BLM)
is usually a good way to absorb the dominant part of two- and
higher-order contributions~\cite{Brodsky:1983gc,LM93}.
On an anisotropic lattice, it may make sense to define separate scales
for temporal and spatial gluons.
These results are of interest in any case:
even if anisotropic lattice calculations require a non-relativistic
interpretation for heavy quarks, anisotropy remains a useful tool for
improving the signal-to-noise ratio.

Finally, after these problems are resolved, it will be important to 
combine the results with numerical simulation data to obtain the
matrix elements relevant to experimental measurements of charmed
hadrons.

\section*{Acknowledgments}

The authors would like to thank Shoji Hashimoto and Masataka Okamoto
for useful discussions.
A.S.K.\ would like to thank Akira Ukawa and the Center for Computational
Physics for hospitality while this work was being completed.
H.M. is supported by the center-of-excellence (COE) program at
Research Center for Nuclear Physics, Osaka University.
T.O. is supported by the Grant-in-Aid of the Ministry
of Education (No.\ 12640279).
Fermilab is operated by Universities Research Association Inc.,
under contract with the U.S.\ Department of Energy.

\appendix

\section{Feynman rules}
\label{sec:Feynman}
\begin{figure}[bp]
\begin{center}
\leavevmode\psfig{file=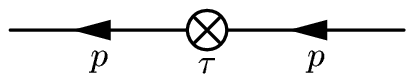,width=0.40\figwidthb} \hspace{2.3cm}
\leavevmode\psfig{file=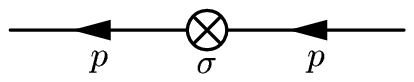,width=0.40\figwidthb} \vspace{-0.6cm}\\
(a) \ ${\cal V}^{\rm (MF)}_4(p)$ \hspace{3.7cm}
(b) \ ${\cal V}^{\rm (MF)}_{\sigma}(p)$ \\
\end{center}
\caption{Feynman rules required for the one-loop perturbation theory.}
\label{fig:Feynman}
\end{figure}

The Feynman rules for perturbative calculation are shown 
the same as in Ref.~\cite{MKE98} except for two points.
One is that the $c_B$ is replaced by $r_s c_B$ as was mentioned
in Sec.~\ref{sec:qaction} and the other is the gluon propagator.
With a gauge fixing term that is symmetric under exchange of
temporal and spatial axis, the free propagator of the gauge field, in
Feynman gauge, is
\begin{equation}
G_{\mu\nu}(k) =
 \left\{
   \begin{array}{ll}
   G_{\tau}(k)~~  & \mu=\nu=4 \\
   G_{\sigma}(k)  & \mu=\nu < 4 \\
   0              & \mu \neq \nu
 \end{array}
\right.
\end{equation}
\begin{eqnarray}
 G_{\sigma} &=& \frac{\delta_{ab} \xi}
       { \hat{k}_i^2 + \xi^2 \hat{k}_4^2 + \xi^2 \lambda^2},
  \nonumber \\
 G_{\tau} &=& \frac{1}{\xi^2} G_{\sigma}(k).
\end{eqnarray}
where we replaced the anisotropy parameter $\gamma_G$ with
the tree-level value~$\xi$.
A~fictitious gluon mass~$\lambda$ is introduced to regulate
infrared divergences.

The Feynman diagram for the counter-term from the mean field
is obtained by expanding
\begin{eqnarray}
 u_{\tau}    &=&  1 + g^2 u_{\tau}^{[1]}   + O(g^4) ,  \\
 u_{\sigma}  &=&  1 + g^2 u_{\sigma}^{[1]} + O(g^4).
\end{eqnarray}
With the replacement of the link variables as Eq.~(\ref{eq:MFreplace})
in the action (\ref{eq:mass_action}), the Feynman rules required for
the the one-loop calculation are 
\begin{eqnarray}
{\cal V}^{\rm (MF)}_4(p)
  &=&  g^2 u_{\tau}^{[1]} [\gamma_4 i \sin p_4 - \cos p_4 ]
  \\
{\cal V}^{\rm (MF)}_{\sigma}(p)
  &=&  g^2 u_{\sigma}^{[1]} \sum_{j} [ \zeta \gamma_j i \sin p_j
                                   - r \zeta \cos p_j ]
\end{eqnarray}
with the diagrams (a) and (b) in Fig.~\ref{fig:Feynman},
respectively.

\section{Explicit expressions of one-loop corrections}
\label{sec:expressions}

In the following, we show the explicit representations of the
self-energy and the vertex correction.
In order to simplify the expressions, we introduce the following
abbreviations:
\begin{eqnarray} 
	c_{\mu}&=&\cos k_{\mu}, \hspace{2.2cm}
	s_{\mu} = \sin k_{\mu}, \nonumber \\
	\check{c}_{\mu}&=&\cos\left(\half k_{\mu} \right), \hspace{1.6cm}
	\check{s}_{\mu} = \sin\left(\half k_{\mu} \right), \nonumber \\
	C_{4}&=&\cos(iM_{1}+k_4),\hspace{0.9cm} 
    S_{4} = \sin(iM_{1}+k_4), \nonumber \\
	\check{C}_{4}&=&\cos\left(iM_{1}+\half k_4\right), \hspace{0.5cm}
    \check{S}_{4} = \sin\left(iM_{1}+\half k_4\right).
\end{eqnarray} 
\begin{equation}
	\vec{s}^2         = \sum_{i=1}^{3} s_{i} s_{i}, \hspace{1cm}
	\check{\vec{c}}^2 = \sum_{i=1}^{3}\check{c}_i\check{c}_i, \hspace{1cm}
	[\vec{s}\cdot\check{\vec{c}}]  = 
		\sum_{i=1}^{3} s_{i}^2 \check{c}_{i}^2.
\end{equation} 
To reduce the volume of notation, we also define
\begin{equation}
	c'_B = r \zeta c_B, \hspace{1cm}  c'_E = r c_E,
\end{equation}
which always appear in these combinations.
To reduce the Dirac matrix structure, it is convenient to introduce
$G^Q=\pm 1$, with the upper (lower) sign for massive quarks
(anti-quarks) on the external leg, and $G$ and $H$, defined by
\begin{equation}
	G\Gamma = \gamma_4  \Gamma\gamma_4   , \quad
	H\Gamma = \gamma^\mu\Gamma\gamma_\mu ,
	\label{eq:GandH}
\end{equation}
with an implied sum on~$\mu$.
The following quantities are convenient for representing the one-loop
expressions below and in our integration programs:
\begin{eqnarray}
	A^q&=&-i  G  \check{c}_{4} - \check{s}_{4}, \hspace{2.45cm}
	A^Q = -i G^Q \check{C}_{4} - \check{S}_{4}, \\
	B^q&=&-i  G  {s}_{4} + 2 \check{s}_4^2 +
		2 r \zeta \check{\vec{s}}^2, \hspace{0.8cm}
	B^Q = -i G^Q {S}_{4} + 1 + m_{0} - C_4 +
		2 r \zeta \check{\vec{s}}^2,     \\
	E^q&=&-i \zeta + \half c'_{E} G  s_4, \hspace{1.9cm}
	E^Q = -i \zeta + \half c'_{E}G^Q S_4, \\
	J^q&=&-i G c_4 + s_4 ,\hspace{2.4cm}
	J^Q = -i G^Q C_4 + S_4,
\end{eqnarray}
and
\begin{eqnarray}
	\bar{B}^q&=&i  G  {s}_{4} + 2 \check{s}_4^2 +
		2 r \zeta \check{\vec{s}}^2, \hspace{0.8cm}
	\bar{B}^Q = i G^Q {S}_{4} + 1 + m_{0} - C_4 + 
		2 r \zeta \check{\vec{s}}^2, \\
	\bar{J}^q&=&i  G  c_4 + s_4, \hspace{2.4cm}
	\bar{J}^Q = i G^Q C_4 + S_4. 
\end{eqnarray}
The symbols with superscript $q$ are essentially massless versions of 
those with superscript~$Q$.

From the Feynman rules in Appendix~\ref{sec:Feynman}, the contributions
to the self-energy from the rainbow diagram, Fig.~\ref{fig:diagram}(a),
are
\begin{eqnarray} 
\Sigma_a(iM_1,{\bf 0})
  &=&  g^2_{\sigma} \Sigma_{a\sigma}^{[1]}(iM_1,{\bf 0})
     + g^2_{\tau}   \Sigma_{a\tau}^{[1]}(iM_1,{\bf 0})  + O(g^4), \\
 \Sigma_{a\sigma}^{[1]}(iM_1,{\bf 0})
 &=& C_F \int \frac{d^4 k}{(2\pi)^4}
     S_Q(iM_1+k_4,\vec{k}) G_{\sigma}(k) 
    \left[\check{\vec{c}}^2 \bar{B}^Q(E^Q)^2 +r^2\zeta^2\check{\vec{s}}^2 B^Q 
     \right. \nonumber \\
 & & \hspace{1.5cm}   \left.
     + i r \zeta^2\vec{s}^2 E^Q
     + \left( i \zeta c'_B E^Q + \quarter c_B^{\prime 2} B^Q \right)
           \left(\vec{s}^2\check{\vec{c}}^2-[\vec{s}\cdot\check{\vec{c}}]
      \right) \right],   \\
\Sigma_{a\tau}^{[1]}(iM_1,{\bf 0})
  &=& C_F \int \frac{d^4k}{(2\pi)^4} S_Q(iM_1+k_4,\vec{k}) G_{\tau}(k)
    \nonumber \\
  &  & \hspace{2cm} \times
     \left[ (A^{Q})^2B^Q+\vec{s}^2\left(i\zeta c'_E \check{c}_4 A^Q G^Q
         +\quarter c_E^{\prime 2} \check{c}_4^2 \bar{B}^Q \right)
     \right]. 
\end{eqnarray} 
where
\begin{equation}
S_Q(p) =  \left[ \sin^2 p_4  +\zeta^2 \sum_j \sin^2 p_j 
              +\left\{ \half  (\hat{p}_{4}^2
              + r \zeta \hat{\vec{p}}^2) + m_{0} \right\}^2\right]^{-1}.
\end{equation}
Similarly, the contributions from the tadpole diagram, 
Fig.~\ref{fig:diagram}(b), are
\begin{eqnarray} 
 \Sigma_b(iM_1,{\bf 0}) 
  &=&   g^2_{\sigma} \Sigma_{b\sigma}^{[1]}(iM_1,{\bf 0}) 
      + g^2_{\tau}   \Sigma_{b\tau}^{[1]}(iM_1,{\bf 0})  + O(g^4),  \\
 \Sigma_{b\sigma}^{[1]}(iM_1,{\bf 0})
 &=& - \half  C_F \int \frac{d^4 k}{(2\pi)^4} 
       G_{\sigma}(k) 3r\zeta,  \\
 \Sigma_{b\tau}^{[1]}( iM_1,\veg{0} ) 
 &=& - \half  C_F \int \frac{d^4 k}{(2\pi)^4} 
           G_{\tau}(k) \left( G^Q\sinh M_1+\cosh M_1 \right).
\end{eqnarray} 
The derivative of the self-energy with respective to $p_4$ is 
separated into temporal and spatial contributions
\begin{equation} 
 \dot{\Sigma}(iM_1,{\bf 0})
 =    g^2_{\sigma} \dot{\Sigma}_{\sigma}^{[1]}(iM_1,\veg{0})
   +  g^2_{\tau}   \dot{\Sigma}_{\tau}^{[1]}(iM_1,\veg{0})
   + O(g^4)
\end{equation} 
The contributions from Fig.~\ref{fig:diagram}(a) are
\begin{eqnarray} 
 \dot{\Sigma}_{a\sigma}^{[1]}(iM_1,\veg{0})
 &=& -iC_F \int \frac{d^4 k}{(2\pi)^4} G_{\sigma}(k) 
    \left[ -2 S_Q(iM_1+k_4,\vec{k})^2 S_4
     \left(1 + 2 r \zeta \check{\vec{s}}^2 + m_{0}\right)  \right.\\
 & & \hspace{-1cm} \times  \left\{ \check{\vec{c}}^2 (E^Q)^2 \bar{B}^Q
       +r^2\zeta^2\check{\vec{s}}^2 B^Q 
       + i r \zeta^2 \vec{s}^2 E^Q
     +\left( i \zeta  c'_B E^Q + \quarter  c_B^{\prime 2} B^Q \right)
      \left( \vec{s}^2 \check{\vec{c}}^2 - [\vec{s}\cdot\check{\vec{c}}]
      \right)   \right\}    \nonumber \\
 & & \hspace{-1cm} \left.
      +  S_Q(iM_1+k_4,\vec{k}) 
    \left\{  \check{\vec{c}}^2 (E^Q)^2  \bar{J}^Q
            +r^2 \zeta^2 \check{\vec{s}}^2 J^Q 
           +\quarter c_B^{\prime 2} (\vec{s}^2\check{\vec{c}}^2
                              - [\vec{s}\cdot\check{\vec{c}}])J^Q
      \right\} \right] \nonumber \\
 \dot{\Sigma}_{a\tau}^{[1]}(iM_1,\veg{0})
 &=& -iC_F \int \frac{d^4 k}{(2\pi)^4} G_{\tau}(k)
    \left[ -2 S_Q(iM_1+k_4,\vec{k})^2 S_4
     \left(1 + 2 r \zeta \check{\vec{s}}^2 + m_{0}\right) \right.\\
 & &  \hspace{-1cm} \times 
     \left\{  (A^Q)^2B^Q+\vec{s}^2\left(i\zeta c'_E \check{c}_4A^QG^Q
     +\quarter c_E^{\prime 2} \check{c}_4^2 \bar{B}^Q \right)
 \right\}
    \nonumber \\
 & &  \hspace{-1cm}
    +S_Q(iM_1+k_4,\vec{k}) \left. \left\{  -2i(A^Q)^2 B^Q G^Q
          +\zeta c'_E\check{c}_4\vec{s}^2 A^Q  
       +\quarter c_E^{\prime 2}\check{c}_4^2\vec{s}^2\bar{J}^Q
          +(A^Q)^2 J^Q \right\} \right]. \nonumber 
\end{eqnarray}
There is only one contribution from Fig.~\ref{fig:diagram}(b)
\begin{equation} 
 \dot{\Sigma}_{b}^{[1]}(iM_1,{\bf 0})
 = -\ihalf C_F \int \frac{d^4 k}{(2\pi)^4} G_{\tau}(k)[G^Q 
                                \cosh M_1 +\sinh M_1 ],
\end{equation}
with a temporal gluon.

The vertex function is split as follows:
\begin{equation}
	g^2\Lambda_\Gamma^{[1]} = 
		g^2_{\sigma} \Lambda^{[1]}_{\Gamma \sigma} +
		g^2_{\tau}   \Lambda^{[1]}_{\Gamma \tau}   ,
\end{equation}
and the contributions are
\begin{eqnarray}
	\Lambda^{[1]}_{\Gamma \sigma} &=& C_F \int \frac{d^4 k}{(2\pi)^4} 
		S_Q(iM_1+k_4,\vec{k}) S_q(k) G_{\sigma}(k)  
 \left[
   \frac{1}{3}\check{\vec{c}}^2(H-G)E^q\overline{B}^qE^Q\overline{B}^Q
   \right. \nonumber \\
&& \hspace{0.8cm} 
   +\,\frac{i}{6}\zeta(H-G)
   \left\{ c'_B(\vec{s}^2\check{\vec{c}}^2-[\vec{s}\cdot\check{\vec{c}}])
   +r\zeta\vec{s}^2\right\} (E^q\overline{B}^q+\overline{B}^QE^Q)
   \nonumber\\
&& \hspace{0.8cm}
   +\,\ihalf \zeta
   \left\{ \frac{1}{6}c'_B
   (\vec{s}^2\check{\vec{c}}^2-[\vec{s}\cdot\check{\vec{c}}])
   \left( (H-G)^2-3 \right)+r\zeta\vec{s}^2\right\} (E^q B^Q+B^q E^Q)
 \nonumber \\
&& \hspace{0.8cm}
   +\,\left\{\frac{1}{24}c_B^{\prime 2}
   (\vec{s}^2\check{\vec{c}}^2-[\vec{s}\cdot\check{\vec{c}}])
   \Big((H-G)^2-3\Big)
   +r^2\zeta^2\check{\vec{s}}^2\right\} B^q B^Q \nonumber\\
&& \hspace{0.8cm}
   -\,\zeta^2 
   \left\{
   \frac{1}{6}(\vec{s}^2\check{\vec{c}}^2
              - [\vec{s}\cdot\check{\vec{c}}])\left( (H-G)^2-3 \right)
   +[\vec{s}\cdot\check{\vec{c}}]
   \right\} E^q E^Q \nonumber \\
&& \hspace{0.8cm}
 \left.
   -\,\frac{1}{12}c_B^{\prime 2}\zeta^2\vec{s}^2
   (\vec{s}^2\check{\vec{c}}^2-[\vec{s}\cdot\check{\vec{c}}])(H-G)
   -\frac{1}{3}r^2\zeta^4\vec{s}^2\check{\vec{s}}^2(H-G) \right]
 \Gamma,  \\
	\Lambda^{[1]}_{\Gamma \tau} &=& C_F\int \frac{d^4 k}{(2\pi)^4}
		S_Q(iM_1+k_4,\vec{k}) S_q(k) G_{\tau}(k) \times \nonumber \\
&& \hspace{0.6cm} \left[
   \Big(A^q B^q+\ihalf \zeta c'_E \check{c}_4\vec{s}^2 G\Big)
   \Big(A^Q B^Q+\ihalf \zeta c'_E \check{c}_4\vec{s}^2 G^Q\Big)
  \right.  \nonumber\\
&& \hspace{0.8cm}\left.
    + \frac{1}{3}(H-G)\vec{s}^2
   \Big(i\zeta A^q+\half c'_E \check{c}_4\overline{B}^qG\Big)
   \Big(i\zeta A^Q+\half c'_E
 \check{c}_4\overline{B}^QG^Q\Big)     \right] \Gamma.
\end{eqnarray}

\end{document}